\documentclass[12pt]{iopart}
\usepackage{iopams}
\usepackage{epsfig}   
\usepackage{graphics}
\usepackage[T1]{fontenc}

\begin{document}

\title{Testing the Copernican Principle Via Cosmological Observations} 
\author{Krzysztof Bolejko$^{1,2}$ and J. Stuart B. Wyithe$^{1}$}

\address{$^1$School of Physics, The University of Melbourne, VIC 3010, Australia}
\address{$^2$Nicolaus Copernicus Astronomical Center, Bartycka 18, 00-716 Warsaw, Poland}

\ead{\mailto{bolejko@camk.edu.pl}, \mailto{swyithe@unimelb.edu.au}}

\begin{abstract}
Observations of distances to Type-Ia
supernovae can be explained by cosmological models that include 
either a gigaparsec-scale void, or a cosmic flow, without the need for Dark Energy. 
Instead of invoking dark energy, these
inhomogeneous models instead violate the Copernican Principle.
we show that current cosmological observations 
(Supernovae, Baryon Acoustic Oscillations and 
estimates of the Hubble parameters based on the age of the oldest stars) are not able to rule out inhomogeneous anti-Copernican models.
The next generation of surveys for baryonic acoustic oscillations
will be sufficiently precise to either validate
the Copernican Principle or determine the existence of a local Gpc scale inhomogeneity.
\end{abstract}



\section{Introduction}

The Copernican Principle states that we do not occupy any special place in the Universe. 
The Friedmann--Lema\^itre--Robertson--Walker (FLRW) models are built on this principle, and provide a remarkably precise description of cosmological observations \cite{WMAP5,sdss,2df}.
One could therefore think of this as an indirect demonstration of the Copernican Principle.
However, homogeneous and isotropic FLRW models are not the only solutions of Einsteins equations which are able
 to fit cosmological observations. In particular, a number of inhomogeneous models
 have been  proposed, each of which are able to describe the evolution of distance with 
redshift as measured via type-Ia supernova (SnIa) without need of a cosmological constant
\cite{sn,sncmb,GH08a,GH08b,CFL08}
 (see \cite{C07} for a review). Moreover, these models may be constructed 
in such a way that they describe the details of the cosmic microwave background power spectrum (CMB)~\cite{sncmb,GH08a}.
By eshuing the constraint of inhomogeneity, these inhomogeneous  models violate the Copernican Principle
 and suggest that we live near the center of a large (of several gigaparsecs diameter) highly isotropic void. 

Inhomogeneous models are able to fit a variety of sets of cosmological
observations without containing a cosmological constant because the
lack of homogeneity offers a great degree of flexibility 
\cite{MHE97}. For example, since the last scattering surface is
separated from regions where supernova are observed by great
distances, the property of inhomogeneity allows a model to be
constructed which provides different physical densities in the regions
from which these two sets of observational data are drawn. Thus, what is
required to constrain inhomogeneous solutions to the apparent
acceleration of the Universe includes several sets of data that 
measure a range of observables at comparable  redshifts.
 In this paper we
confront two general classes of inhomogeneous, dark energy free models 
with observations of SnIa, with measurements
of the baryon acoustic oscillations (BAO), and with the variation of the
Hubble parameter with redshift. 
All these observational data are drawn from $z<1.8$.
We begin by describing the Lama\^itre--Tolman model (the
simplest spherically symmetric inhomogeneous generalization of
FLRW model), and then construct a range of 2-parameter models to
describe a local large scale inhomogeneity. We then discuss
constraints on the scale of these inhomogeneities based on existing
and forthcoming data.

\section{The Lema\^itre--Tolman model}\label{LTsec}

The Lema\^itre--Tolman (LT) model \cite{LT} is a spherically symmetric, pressure free, irrotational solution of the Einstein equations.
Its  metric  has the form
\begin{equation}
{\rm d}s^2 =  c^2{\rm d}t^2 - \frac{R'^2(r,t)}{1 + 2 E(r)}\ {\rm
d}r^2 - R^2(t,r) {\rm d} \Omega^2, \label{ds2}
\end{equation}
where $ {\rm d} \Omega^2 = {\rm d}\theta^2 + \sin^2 \theta {\rm
d}\phi^2$. Because of the signature $(+, -, -, -)$, the $E(r)$ function must
obey $E(r) \ge - 1/2.$ Here a prime ($'$) denotes $\partial_r$.

The Einstein equations reduce to the following two
\begin{equation}\label{den}
\kappa \rho(r,t) c^2 = \frac{2M'(r)}{R^2(r,t) R'(r,t)},
\end{equation}
\begin{equation}\label{vel}
\frac{1}{c^2}\dot{R}^2(r,t) = 2E(r) + \frac{2M(r)}{R(r,t)} + \frac{1}{3}
\Lambda R^2(r,t),
\end{equation}
\noindent where $M(r)$ is another arbitrary function and $\kappa = 8 \pi
G/c^4$. Here a dot ($\dot{}$) denotes $\partial_t$.
When $R' = 0$ and $M' \ne 0$, the density becomes infinite. This happens at
shell crossings, and is an additional singularity to the Big Bang that
occurs at $R = 0, M' \neq 0$.  By setting the initial conditions
appropriately the shell crossing singularity can be avoided (see
\cite{HL85} for detail discussion).

Equation (\ref{vel}) can be solved by simple integration:

\begin{equation}\label{evo}
\int\limits_0^R\frac{d\tilde{R}}{\sqrt{2E + \frac{2M}{\tilde{R}} +
\frac{1}{3}\Lambda \tilde{R}^2}} = c \left[t- t_B(r)\right],
\end{equation}
where $t_B$ appears as an integration constant and is an arbitrary function
of $r$. This means that the big bang is not a single event as in the
FLRW models, but occurs at different times at different distances from
the origin.
To define a particular LT model two functions must be specified as initial conditions.
For completeness we note that in the FLRW limit $R(r,t) = r a(t)$ [where $a(t)$ is the scale factor], $M(r)  = M_0 r^3$, and $E(r) = - k_0 r^2$.

\section{Parametrization of Inhomogeneous Models}
\label{TESTsec}

\begin{figure}
\begin{center}
\includegraphics[scale=0.6]{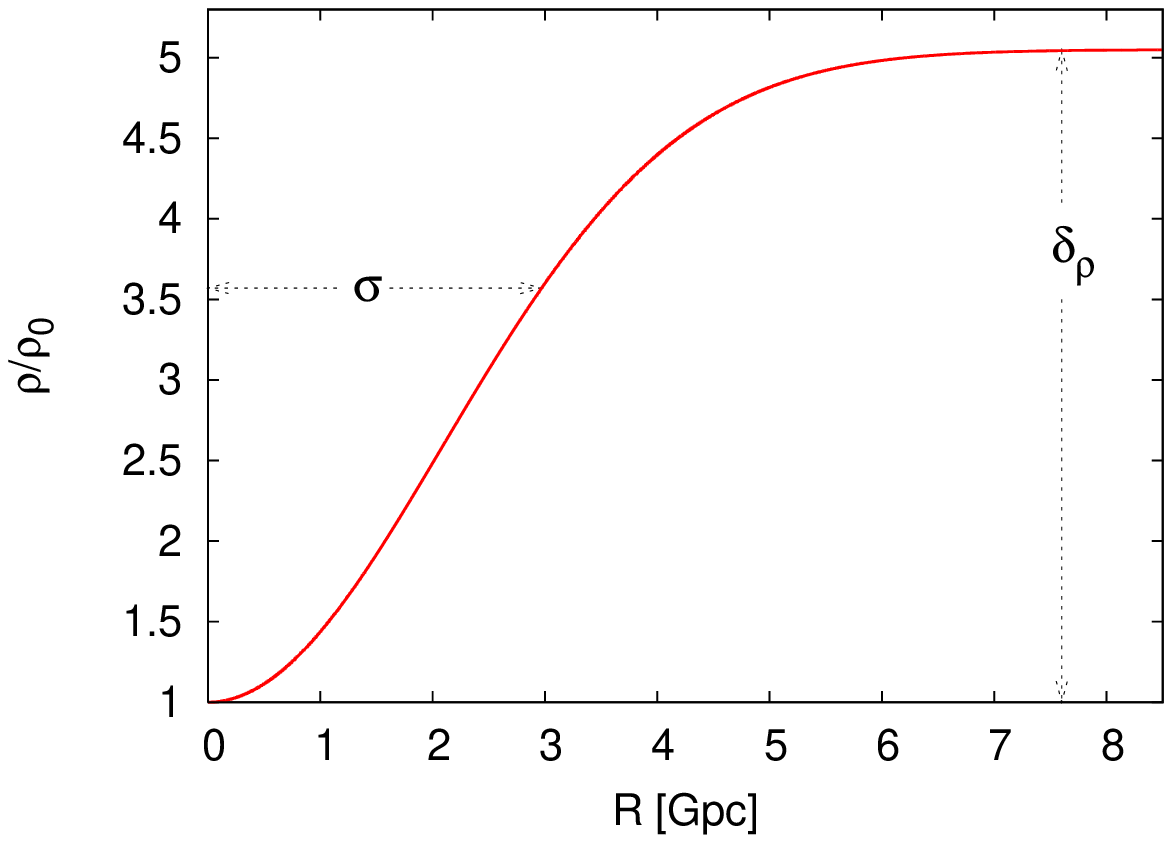}
\includegraphics[scale=0.6]{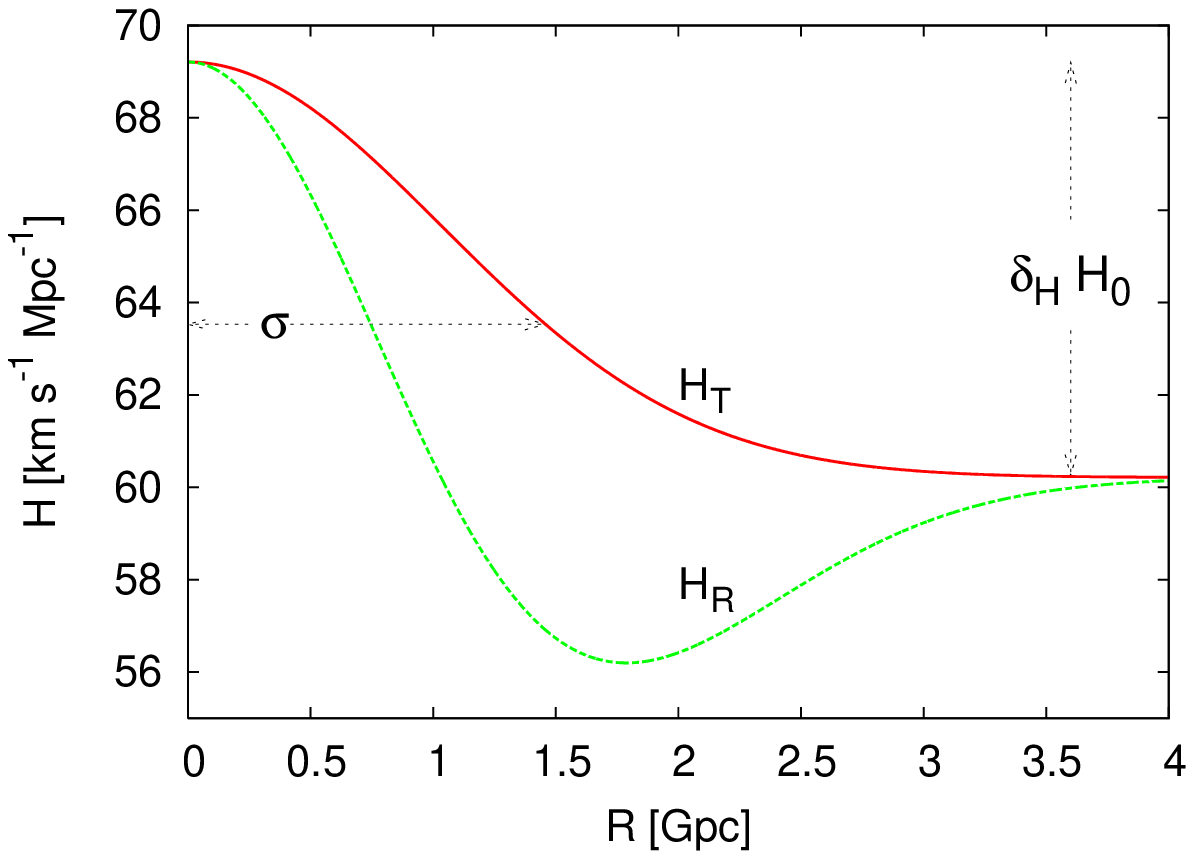}
\caption{{\em Left} panel: The parametrization of the density
distribution within the first family of model at the current instant. 
{\em Right} panel: The parametrization of the expansion rate ($H_T$) at the current instant within the second family of models, as well as the corresponding Hubble parameter
$H_R$.  The plotted profiles were obtained using parameters 
presented in Table \ref{tab1}.}
\label{fig1}
\end{center}
\end{figure}

In this paper we consider two families of models,
in which the cosmological constant is set to zero ($\Lambda = 0$)
and the observer is situated at the origin.
The most popular way to explain SnIa without $\Lambda$ is to postulate a large
 scale void centered near our position \cite{sn,sncmb,GH08a,GH08b,CFL08}.
The density distribution at the current instant in this first family may be parametrized by (see Fig.~\ref{fig1})
\begin{equation}
\rho(t_0,r) = \rho_0 \left[ 1 + \delta_{\rho} - \delta_{\rho} \exp \left( - \frac{r^2}{\sigma^2} \right) \right],
\label{rhofl}
\end{equation}
where $\rho_0 = 0.3 \times  (3H_0^2)/(8\pi G)$.
In these models the big bang is assumed to occur simultaneously at every point ($t_B = 0$).
The functions $M$ and $E$ are then calculated using eqs. (\ref{den}) and (\ref{evo}) respectively.

A second possible explanation for the apparent acceleration observed using SnIa is 
inspired by the so called Hubble Bubble. In the Hubble Bubble phenomenon the expansion 
rate at distances beyond 100 Mpc is postulated to be slower than it is locally \cite{JRK07}.
Such a phenomenon could be described by a second family of models parametrized using (see Fig.~\ref{fig1})
\begin{equation}
H_T(t_0,r) = \frac{\dot{R}}{R} = H_0 \left[ 1 - \delta_{H} + \delta_{H} \exp \left( - \frac{r^2}{\sigma^2} \right) \right].
\label{expfl}
\end{equation}
In these models density is assumed to be homogeneous at the current epoch.
The function $M$ can be calculated using the above relation and eq. (\ref{vel}).
As can be easily checked, these two families of models are regular
at the origin (for origin conditions in the LT models see \cite{MH01}).

In each case the radial geodesics and redshift are calculated from \cite{B47} 
\begin{equation}
 \frac{{\rm d} t}{{\rm d} r} = -  \frac{R'}{\sqrt{1 + 2E}}, \hspace{7mm}\ln (1+z) = \int\limits_0^r {\rm d} \tilde{r} \frac{\dot{R}'}{\sqrt{1+2E}},
\label{nge}
\end{equation}
and the luminosity distance is $D_L = (1+z)^2 R$.

\section{Cosmological Observations}

In this paper we confront the inhomogeneous models described above with
three sets of observations. Firstly, we consider observations of
Type-Ia supernova, which are taken 
from the Union data set \cite{K08}. In addition we also present
the constraints coming from  the
Riess gold data set \cite{R07}, which until recently
has been the most popular sample for testing inhomogeneous models. The second set of cosmological observations 
comprise measurement of the dilation scale of the BAO in the redshift space power-spectrum of 46,748 luminous red galaxies (LRG) from the Sloan Digital Sky Survey
(SDSS). The dilation scale is defined as
\begin{equation}
D_V = \left[ D_A^2 \frac{cz}{H(z)} \right]^{1/3},
\label{dvdf}
\end{equation}
where $D_A$ is the comoving angular diameter distance and $H(z)$ is the Hubble parameter in function of redshift. The measured value of the dilation scale at $z=0.35$ 
is 1370 $\pm$ 64 Mpc \cite{E05}\footnote{It should be noted that the value of 1370 $\pm$ 64 Mpc was obtained
within the framework of the linear perturbations imposed on a homogeneous FLRW 
background. It is still an open question whether such analysis is appropriate
or should be carried out instead in the LT background.
However, we proceed with the value for the $D_V$ provided by Eisenstein et al. 
We will come back to this issue in Sec. \ref{sec6.1} where 
we will try to partially estimate the errors which arise from application of different
background models.}.
The Hubble parameter in equation (\ref{dvdf}) is  related to expansion along the radial direction, and  in the LT model is given by
\begin{equation}
H_R = \frac{\dot{R}'}{R'}.
\label{hrdf}
\end{equation}

In addition to the geometric measurements described above, the Hubble parameter has been estimated as a function of time based on the age of the oldest stars observed in galaxies at different redshifts~\cite{S05}. 
Using eq. (\ref{nge}) it can be shown that within the LT model the Hubble parameter measured this way is also equal to $H_R$.
Given a particular parametrization  the null geodesic equations can be solved to 
calculate the luminosity distance, dilation scale and the Hubble parameter ($H_R$). 
These quantities can then be compared with observations to test the viability of the model by constraining parametrizations via a least square fit.

\section{Testing the Copernican Principle}\label{sec5}

\subsection{Cosmic void models}\label{sec5.1}

\begin{figure}
\begin{center}
\includegraphics[scale=0.37]{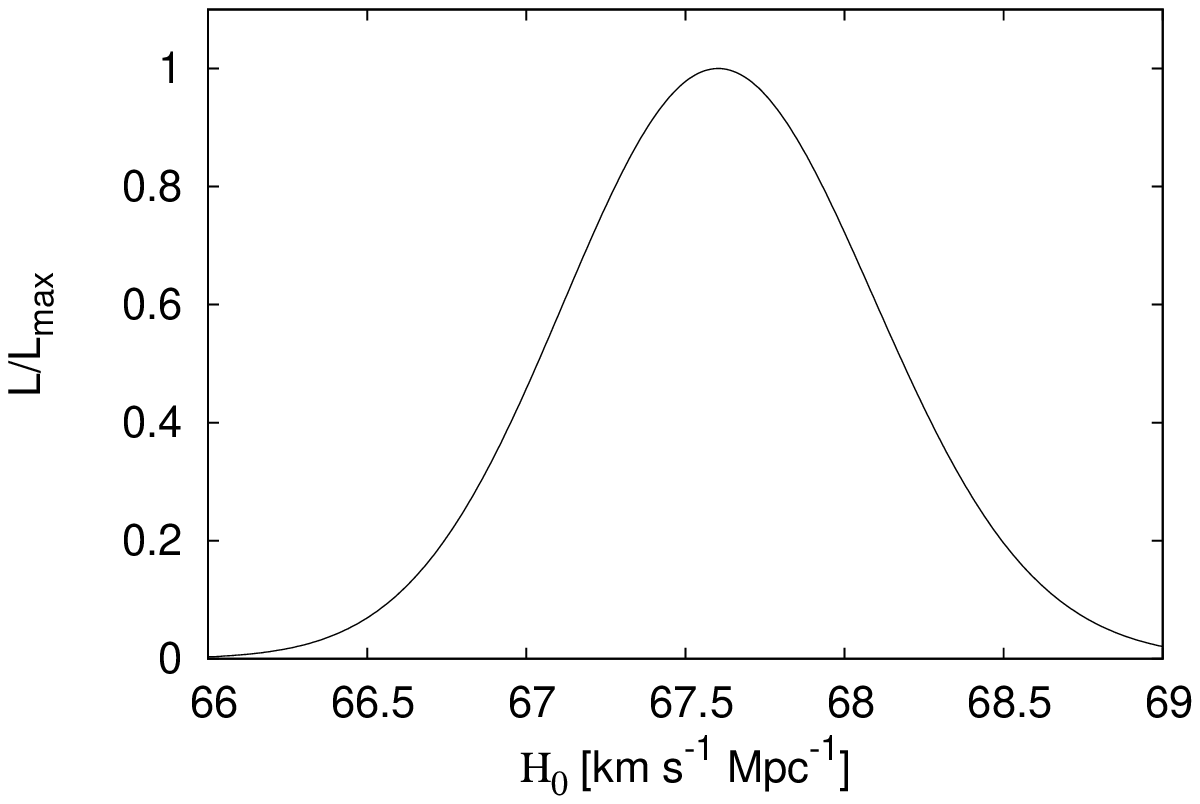}
\includegraphics[scale=0.37]{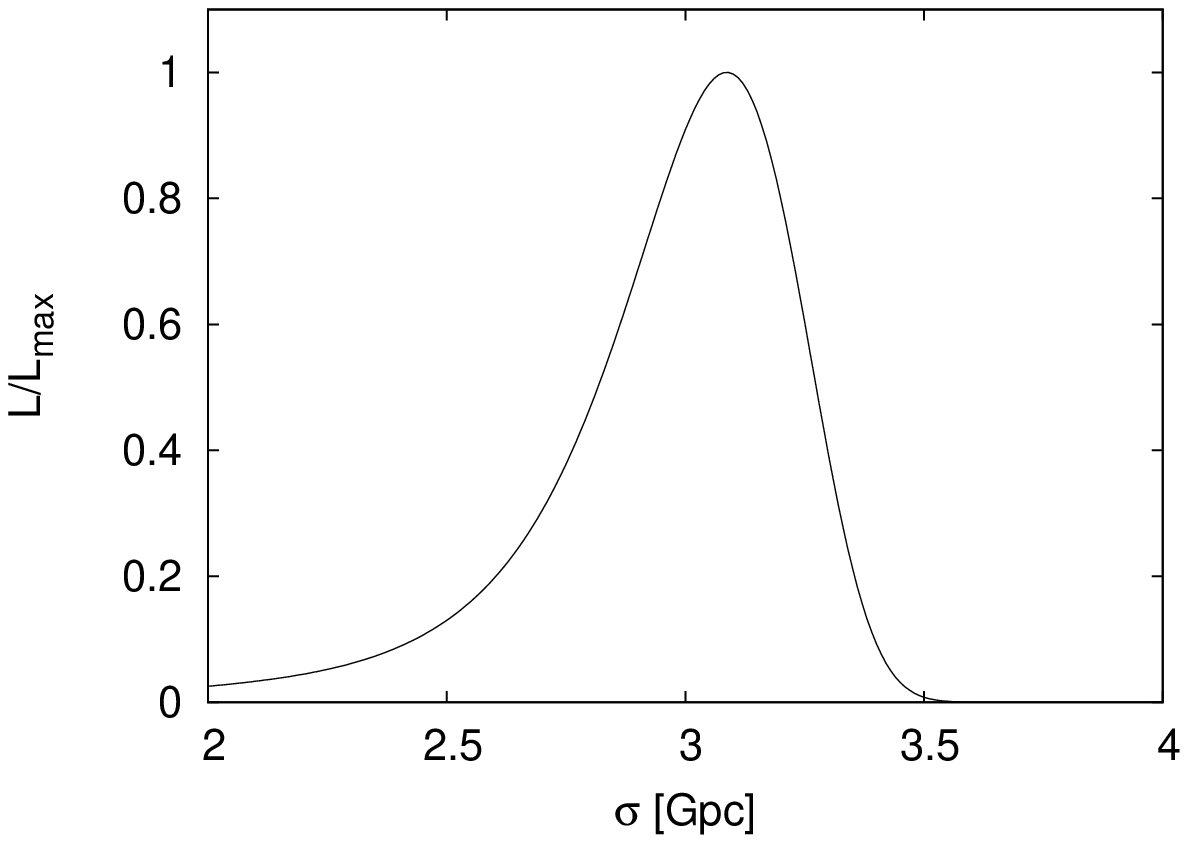}
\includegraphics[scale=0.37]{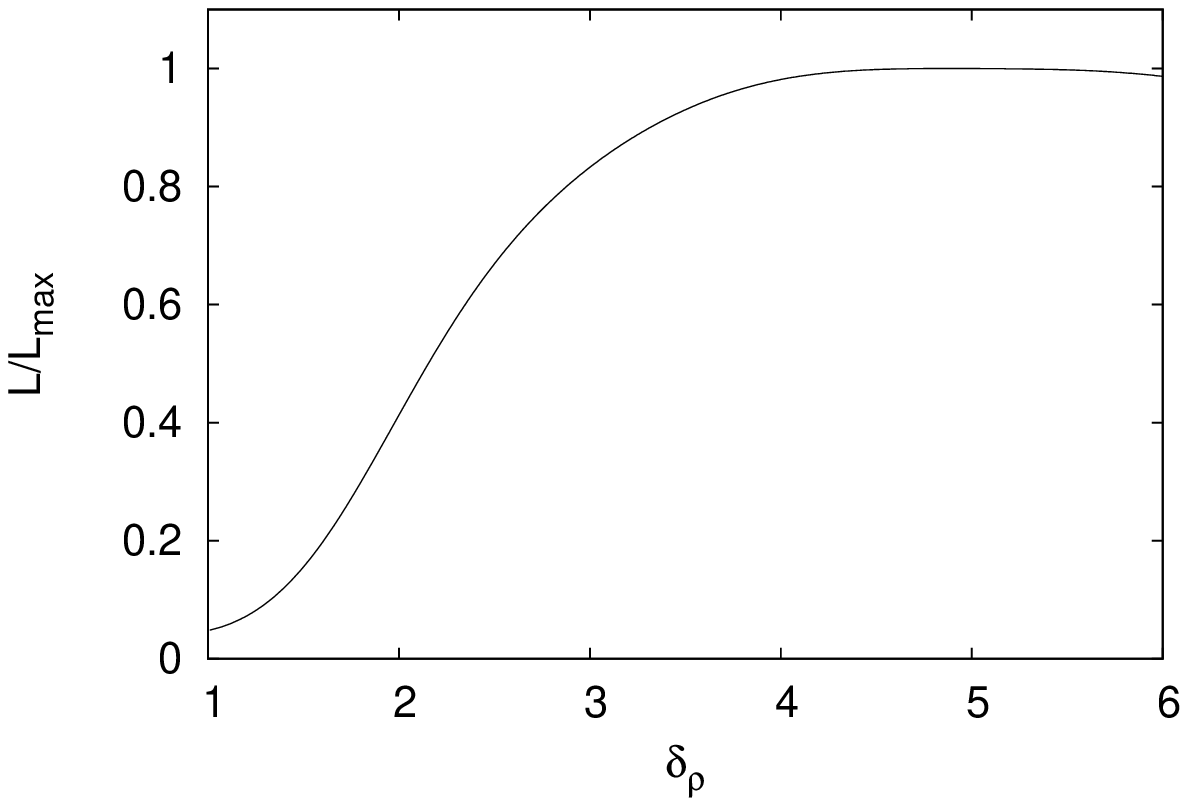}
\caption{Likelihood profiles for $H_0$, $\sigma$, and $\delta_{\rho}$
based on the BAO+H(z)+Union data sets.
{\em Left} panel: The $H_0$ likelihood profile 
obtained after marginalization over $\sigma$ and $\delta_{\rho}$.
{\em Center} panel: The $\sigma$ likelihood profile 
obtained after marginalization over $H_0$ and $\delta_{\rho}$.
{\em Right} panel: The $\delta_{\rho}$ likelihood profile 
obtained after marginalization over $\sigma$ and $H_0$.
The means are presented in Table \ref{tab1}.}
\label{fig2}
\end{center}
\end{figure}

\begin{figure}
\begin{center}
\includegraphics[scale=0.4]{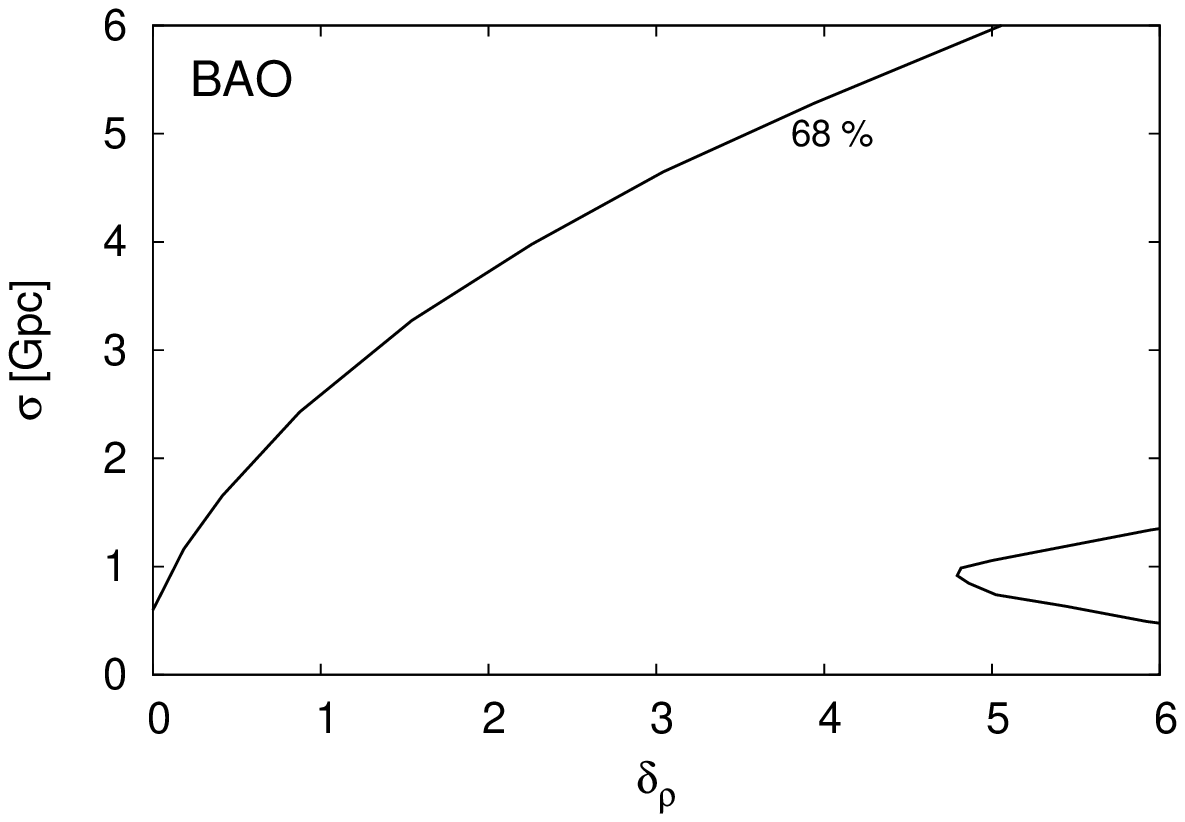}
\includegraphics[scale=0.4]{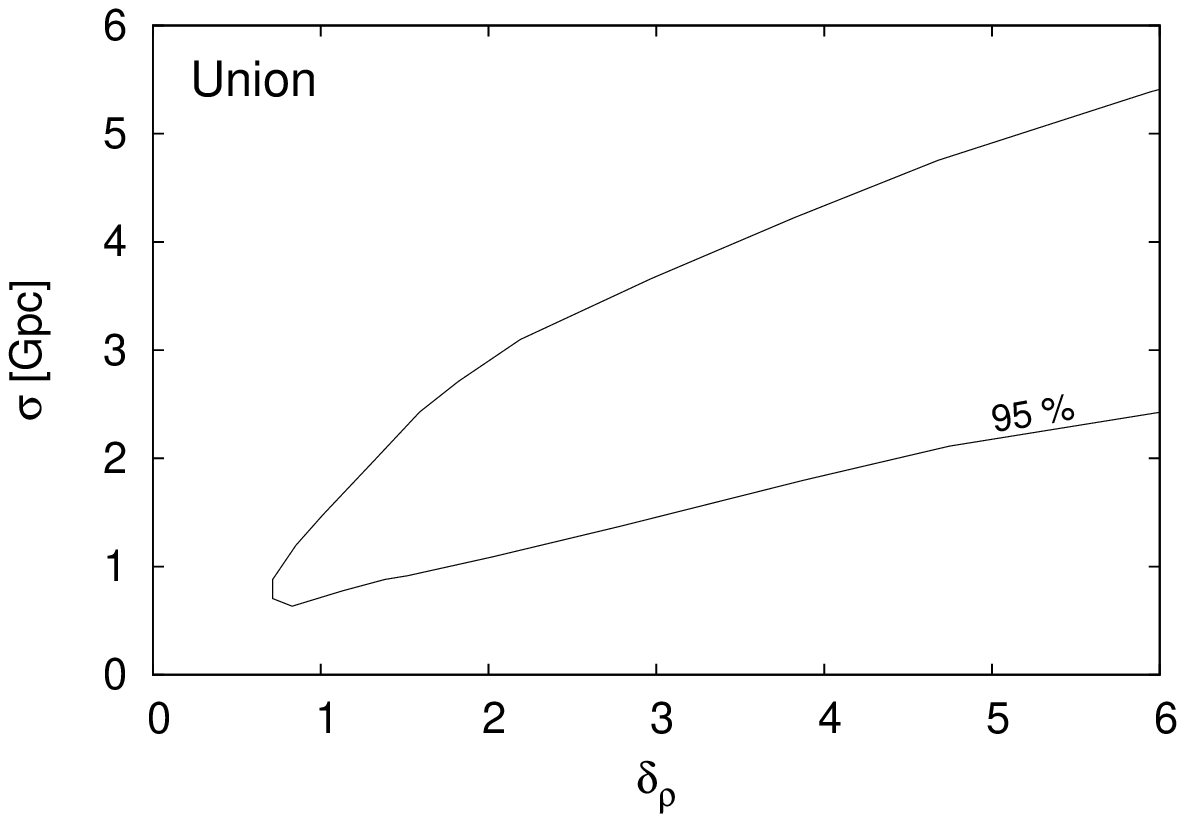}
\includegraphics[scale=0.4]{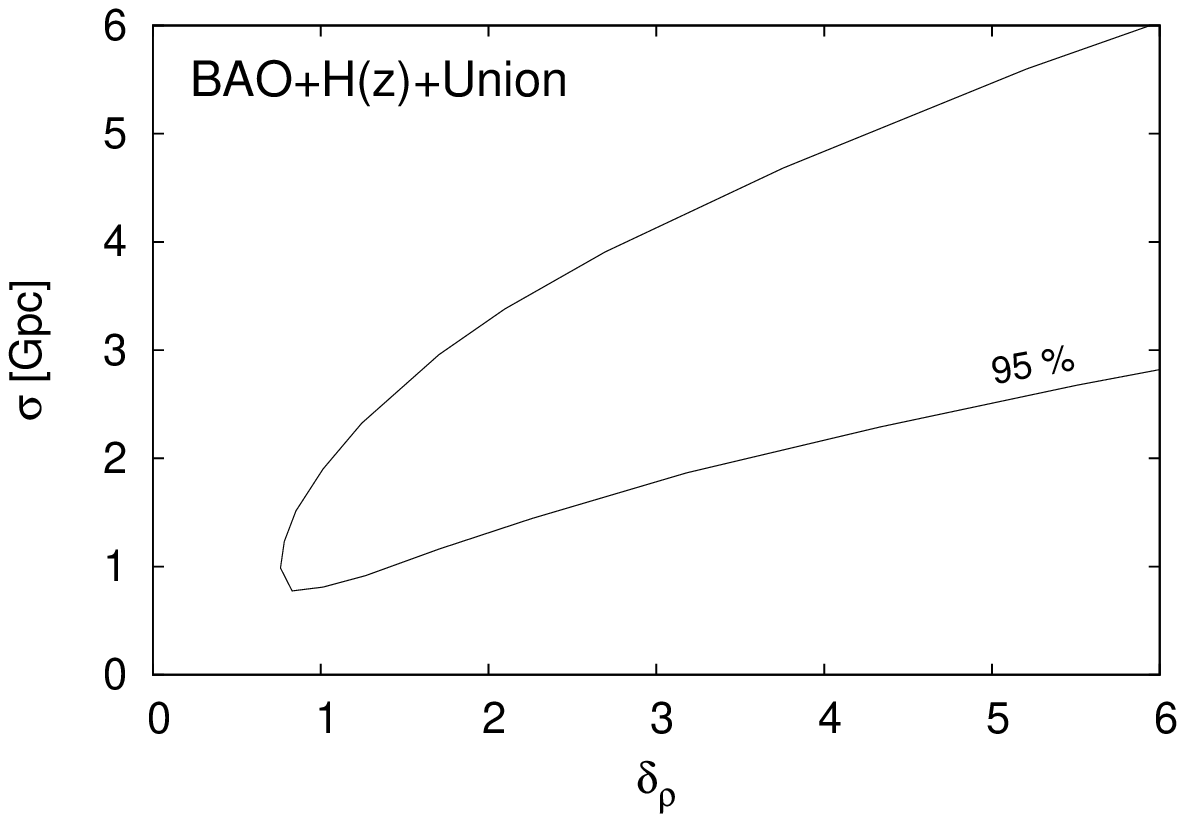}
\includegraphics[scale=0.4]{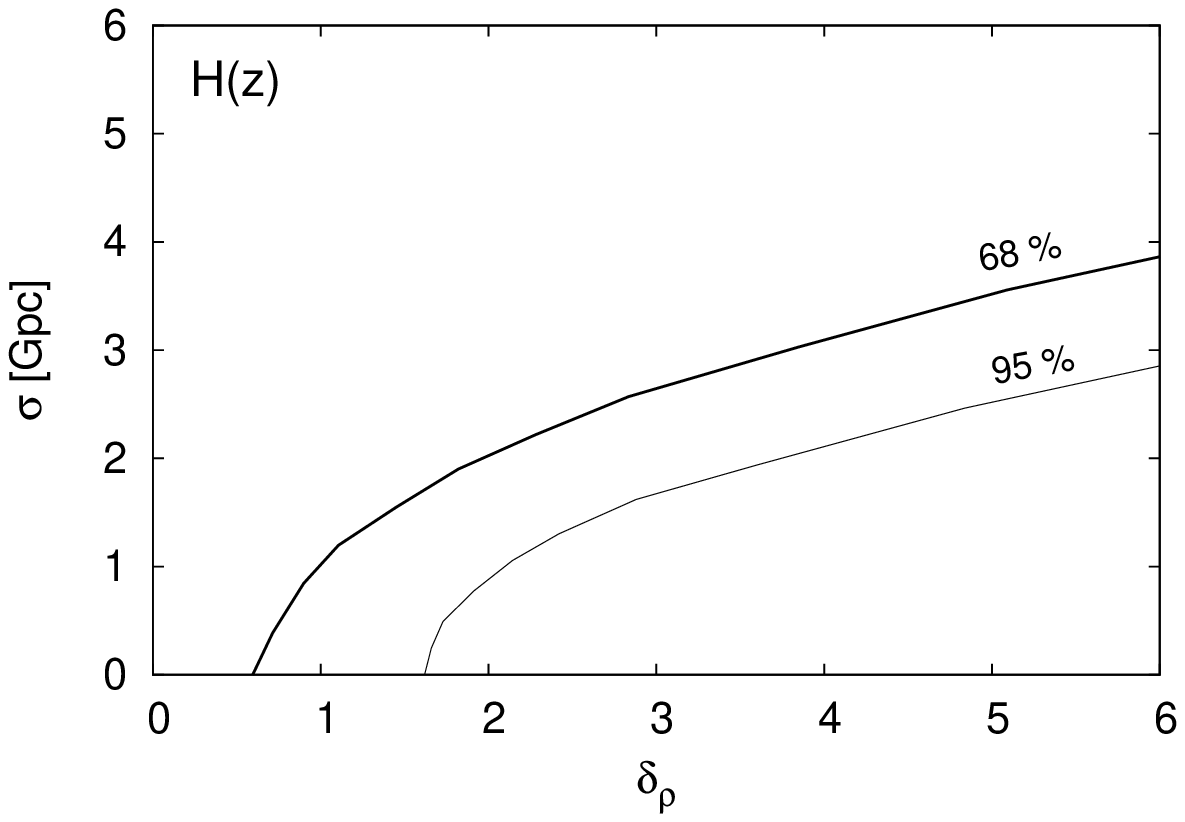}
\includegraphics[scale=0.4]{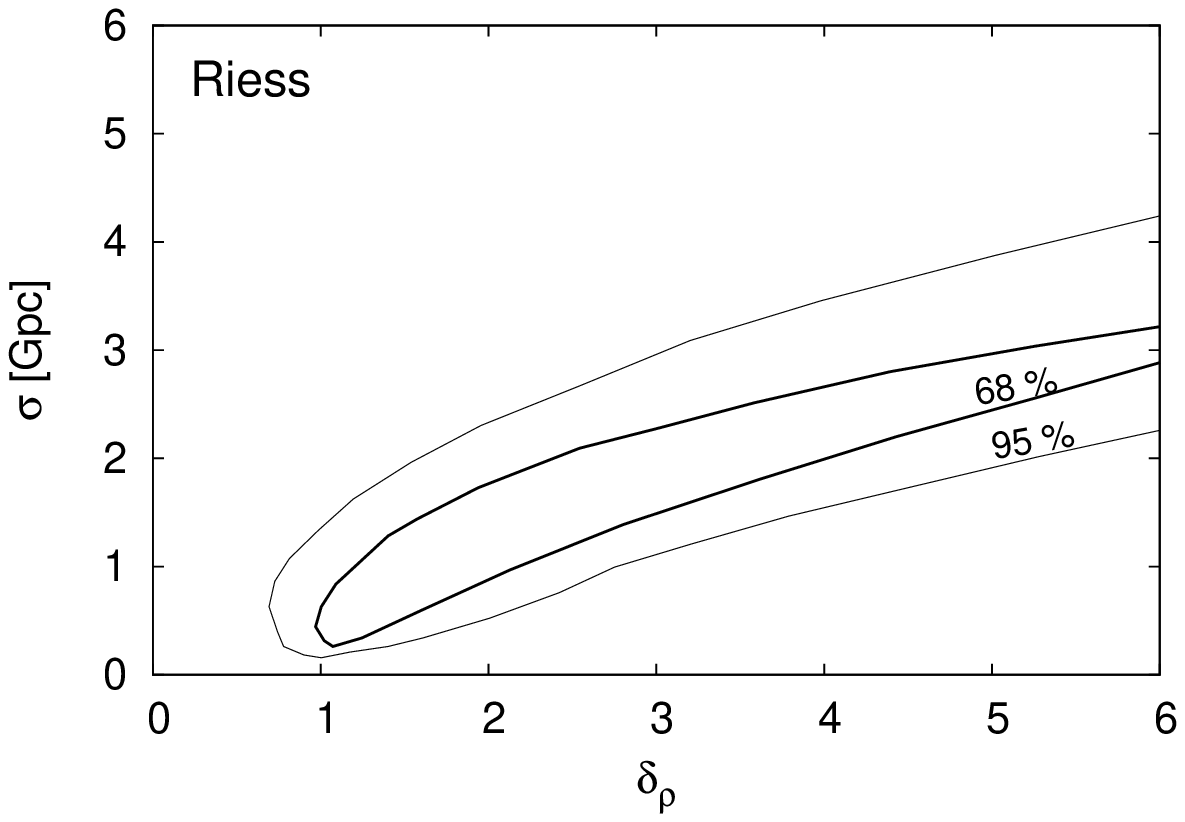}
\includegraphics[scale=0.4]{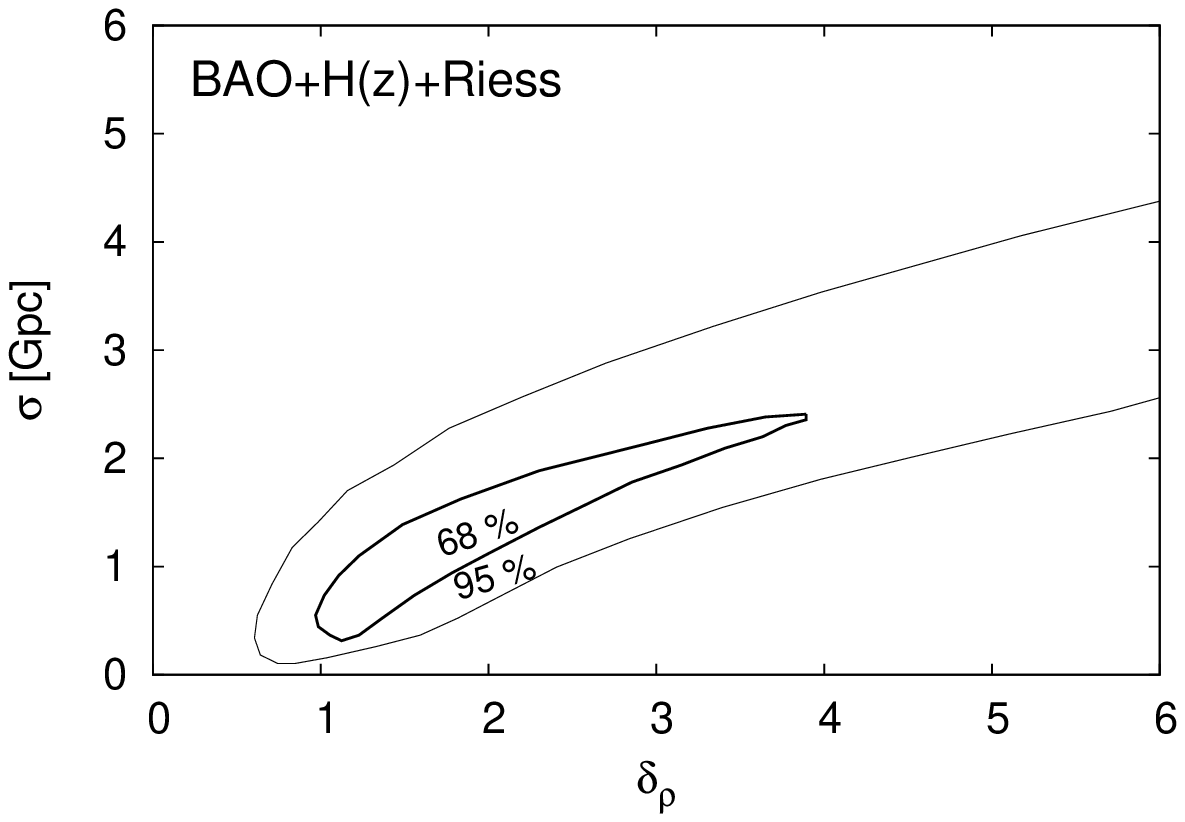}
\caption{
Observational constraints on the parameters of the first family of models (cosmic void models), shown as contour plots of $\chi^2$ for the model given the data,
after marginalizing over $H_0$.
The confidence levels are set assuming a $\chi^2$ distribution.
{\em Upper Left}: constraints from the BAO.
{\em Lower Left}: constraints from H(z).
{\em Upper Center}: constraints from the Union data set.
{\em Lower Center}: constraints from the Riess gold data set.
{\em Upper Right}: joint constraints from 
the BAO+H(z)+Union data sets.
{\em Lower Right}: joint constraints from 
the BAO+H(z)+Riess data sets.}
\label{fig3}
\end{center}
\end{figure}

We begin by constraining the first family  of inhomogeneous models, characterized by $t_B = 0$. 
As can be seen from equation~(\ref{rhofl}), the density increases from $\rho_0$ at the origin to $\rho = (1+\delta_{\rho})\rho_b$ at infinity.
The left panel of Fig.~\ref{fig1} shows an example
profile for which $\delta_{\rho}=4.05$ and $\sigma=2.96$. We note that the profile need not be extrapolated
to infinity. Indeed, owing to the flexibility of inhomogeneous models 
the profile could be modified arbitrarily at larger distances in order to fit other types of observations, including those of the CMB
(see Appendix for more details).
We allow the following three parameters to vary 
within specified ranges  $H_0 \in [64,76]$ km s$^{-1}$ Mpc$^{-1}$ , $\sigma \in [0,6]$ Gpc, and $\delta_{\rho} \in [0,6]$.
The likelihood distributions for these parameters are presented in Fig.~\ref{fig2}. The expected (mean) values derived
from the likelihood profiles are presented in Table \ref{tab1}.
The constraints on the parameters $\delta_{\rho}$ and $\sigma$
are shown as contour plots of $\chi^2$ after marginalizing over $H_0$ in Fig.~\ref{fig3}.
As can be seen the Union data set places tighter constraints on the
models than the Riess data set.
However, it should be emphasized that the 
uncertainties in the 
Union data set used here do not include systematic errors,
which may be present owning to the joining of several supernova surveys. Some of the
systematics are associated with particular surveys, while others are common
to all surveys (intristic variation in supernova explosion mechanism or their possible evolution).
In the standard approach, systematic errors are estimated by adding
a systematic component, $\sigma_{sys}$, in quadrature 
to the statistical error. The
amplitude of $\sigma_{sys}$ is then evaluated  so that the $\chi^2$ per
degree of freedom for the best fitting cosmological model is unity.
However, since in this paper we aim to constrain inhomogeneous
cosmological models, and to determine whether or not they are able
to provide a successful fit to the cosmological observations, 
we have decided not to account for  possible systematical errors.
Indeed, using the procedure described above  
would allow us to fit the supernovae data with almost any model. 
On the other hand, we find that the best-fit model of the family of models studied
in this section (see Table \ref{tab2} for the exact values of model parameters) fits supernova data with $\chi^2 =325.89$ (for 304 degrees of freedom), which means that the model can be ruled out at only the 81.4\% level.
  The overall fit to all three cosmological data sets 
(BAO+H(z)+Union) is 336.55 (for 314 degrees of freedom),
which means that the model can be ruled out at only the 81.7\% level.
These low values of values of $\chi^2$ per degree of freedom imply that 
these observational data cannot rule out anti-Copernican models describing a large cosmic depression.
Within the parametrization assumed, the best fitted void 
has a density contrast of  $\delta_{\rho} \approx 4.5$ and a radius of $\sigma \approx 3.2$ Gpc.

We find that the best constraints come from the supernova data
set. This is mostly because 304 measurements are available, compared
with only 9 for $H(z)$ and just one for BAO. However, as may be seen
from the lower left panel of Fig. 3., the $H(z)$ data prefers large
voids, and so these data also contribute to constraints on
parameter estimates. The weakest constraints come from the BAO
measurement --- see upper left panel of Fig. 3.

\subsection{Cosmic flow models}

\begin{figure}
\begin{center}
\includegraphics[scale=0.37]{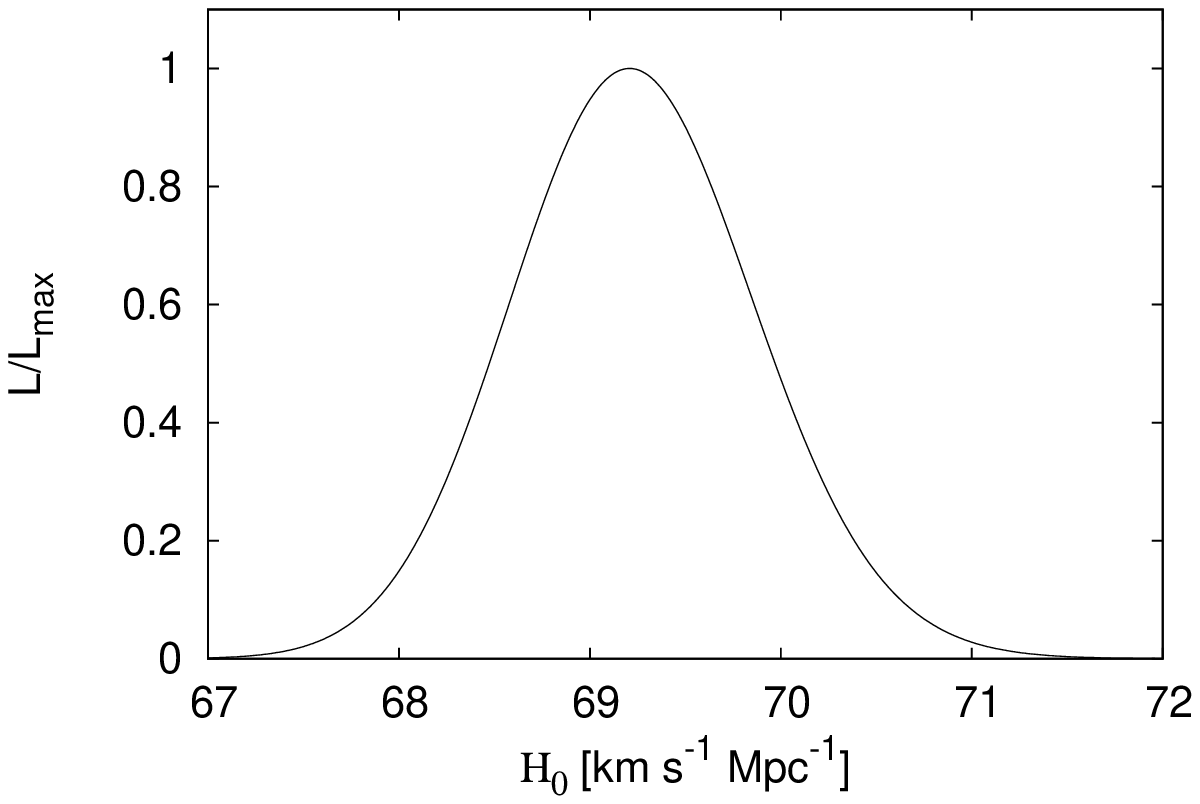}
\includegraphics[scale=0.37]{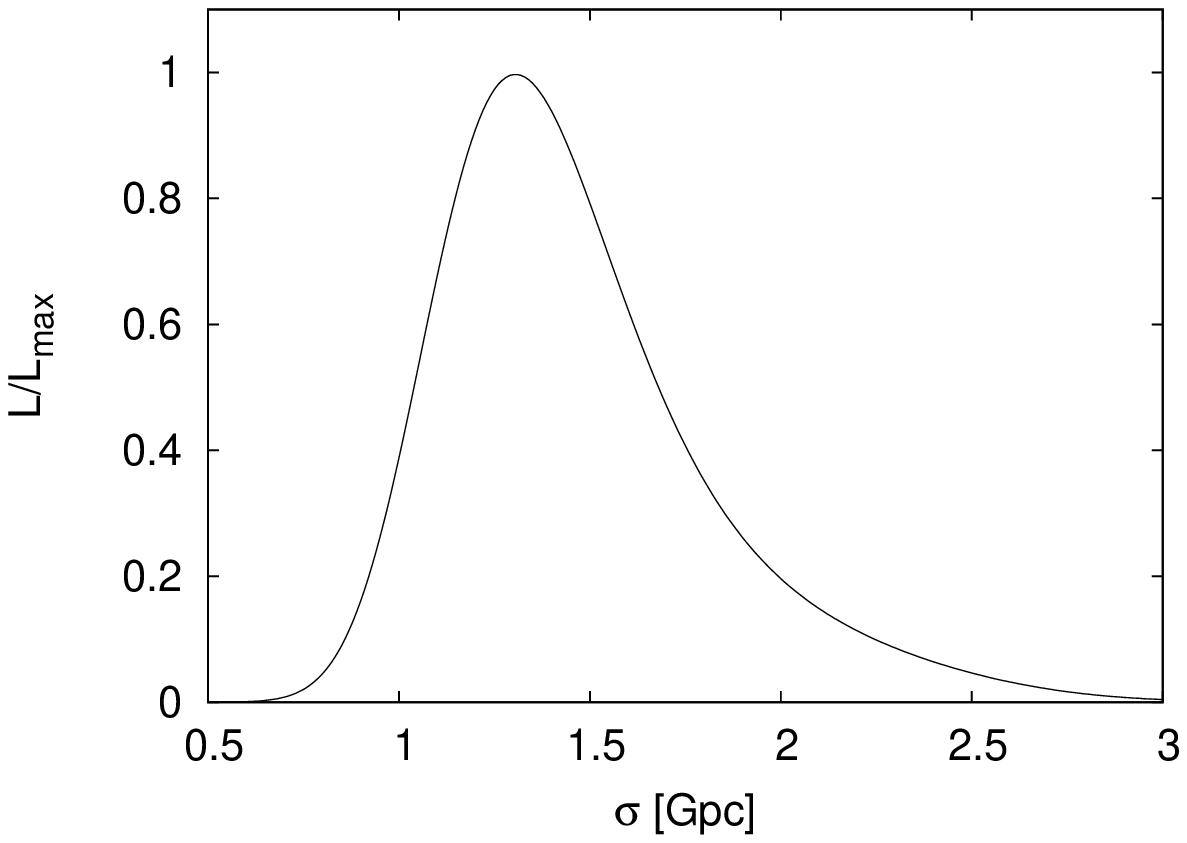}
\includegraphics[scale=0.37]{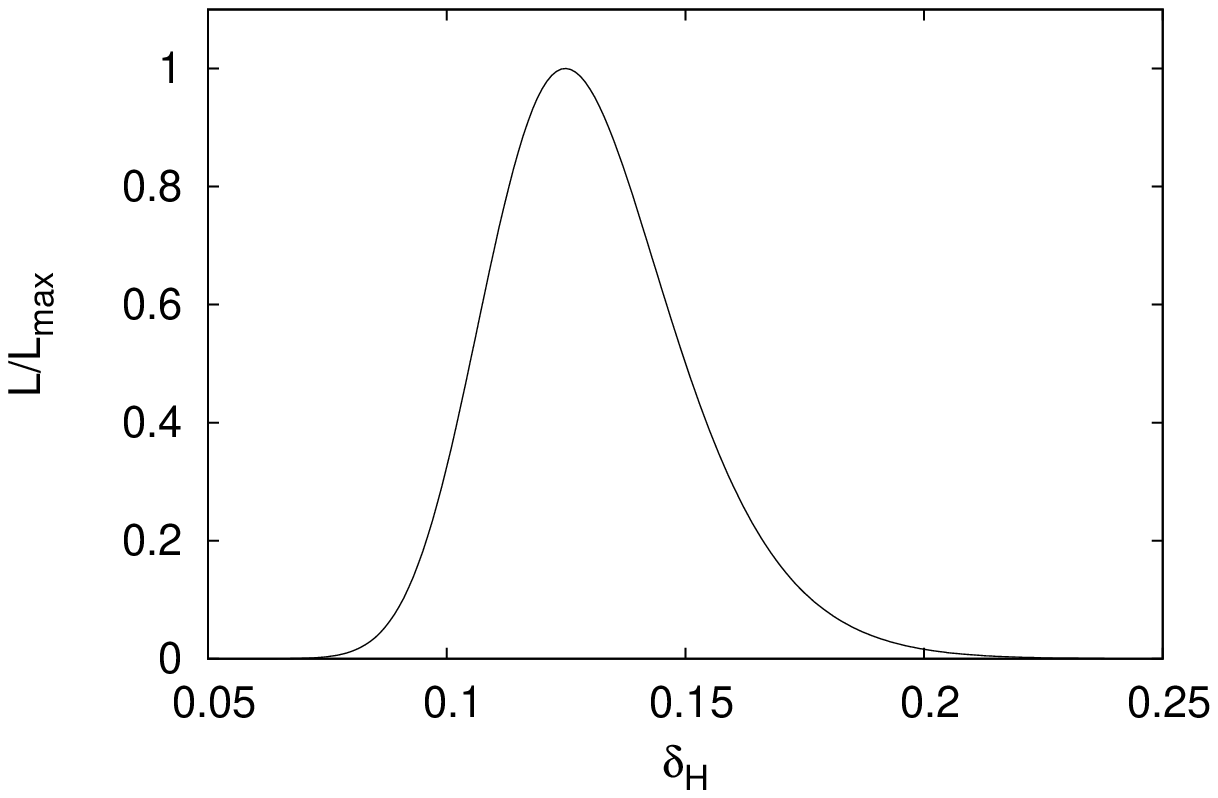}
\caption{Likelihood profiles for $H_0$, $\sigma$, and $\delta_{\rho}$
based on the BAO+H(z)+Union data sets.
{\em Left} panel: The $H_0$ likelihood profile 
obtained after marginalization over $\sigma$ and $\delta_{\rho}$.
{\em Center} panel: The  $\sigma$ likelihood profile 
obtained after marginalization over $H_0$ and $\delta_{\rho}$.
{\em Right} panel: The $\delta_{\rho}$ likelihood profile 
obtained after marginalization over $\sigma$ and $H_0$.
The means are presented in Table \ref{tab1}.}
\label{fig4}
\end{center}
\end{figure}

\begin{figure}
\begin{center}
\includegraphics[scale=0.4]{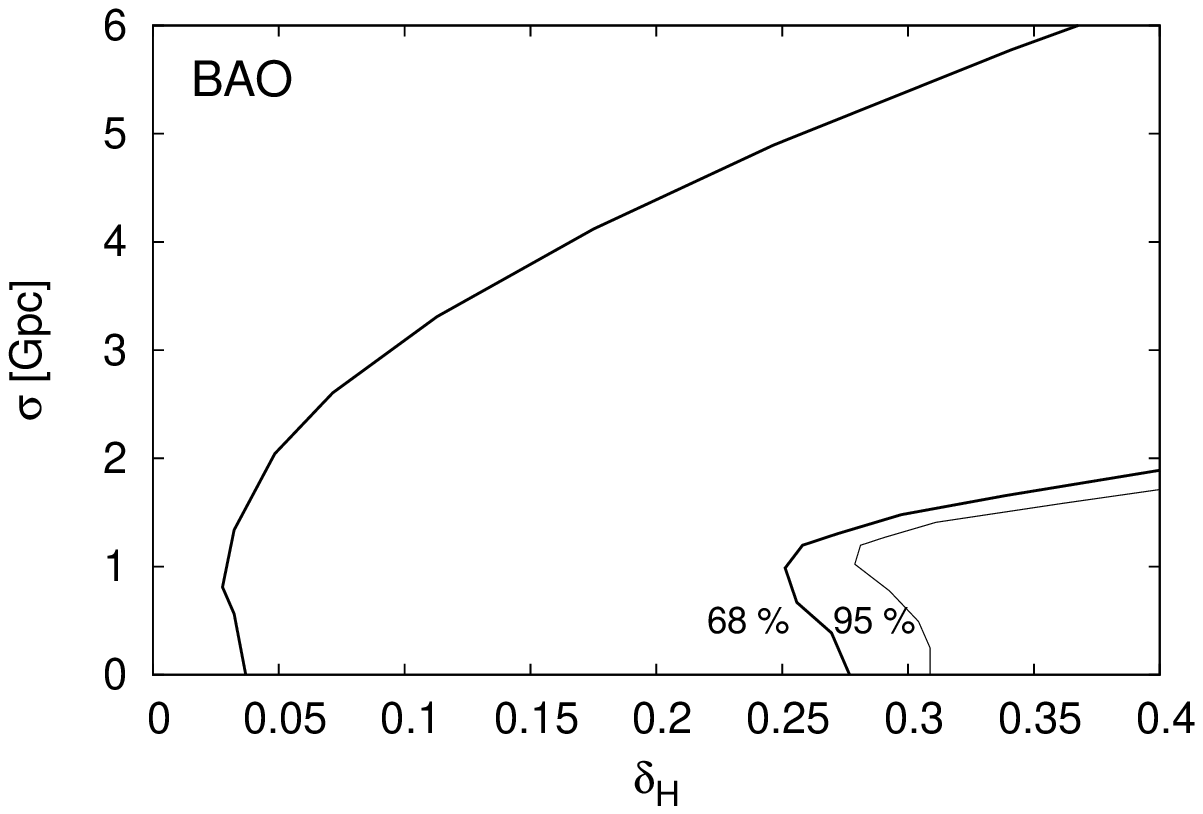}
\includegraphics[scale=0.4]{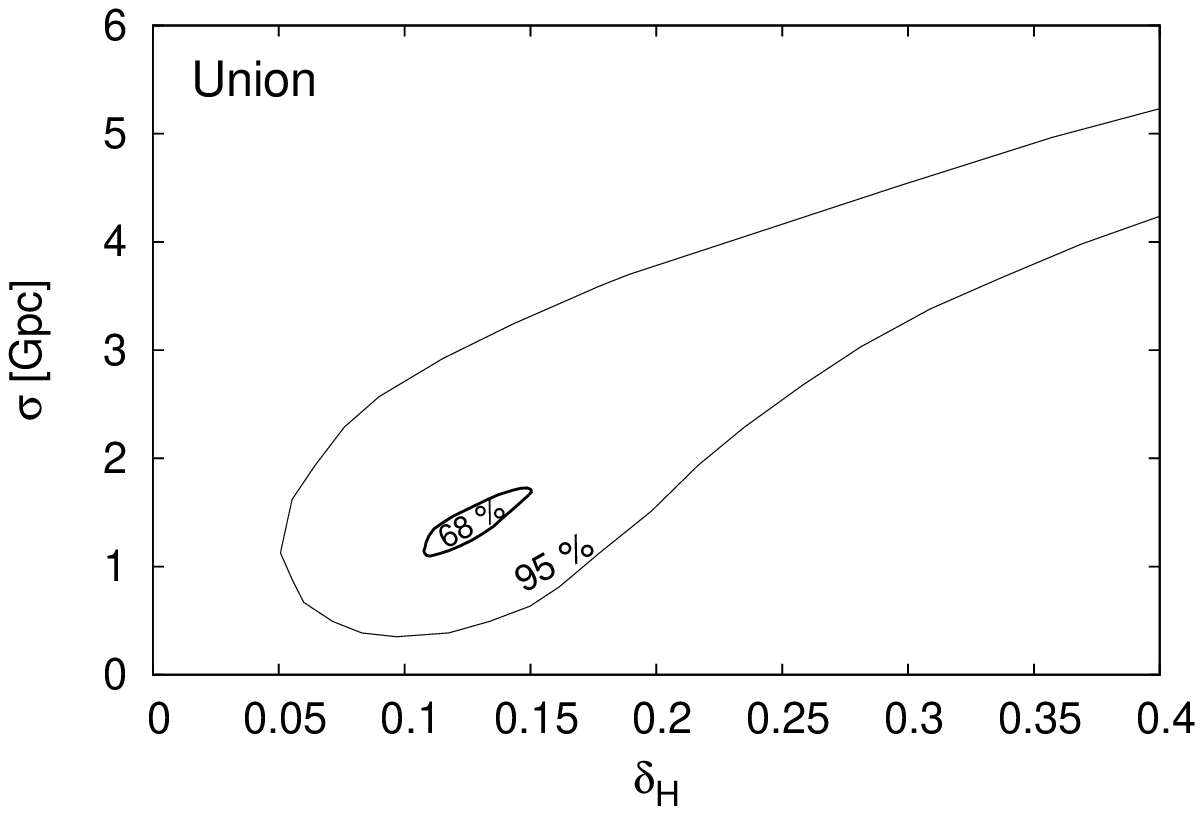}
\includegraphics[scale=0.4]{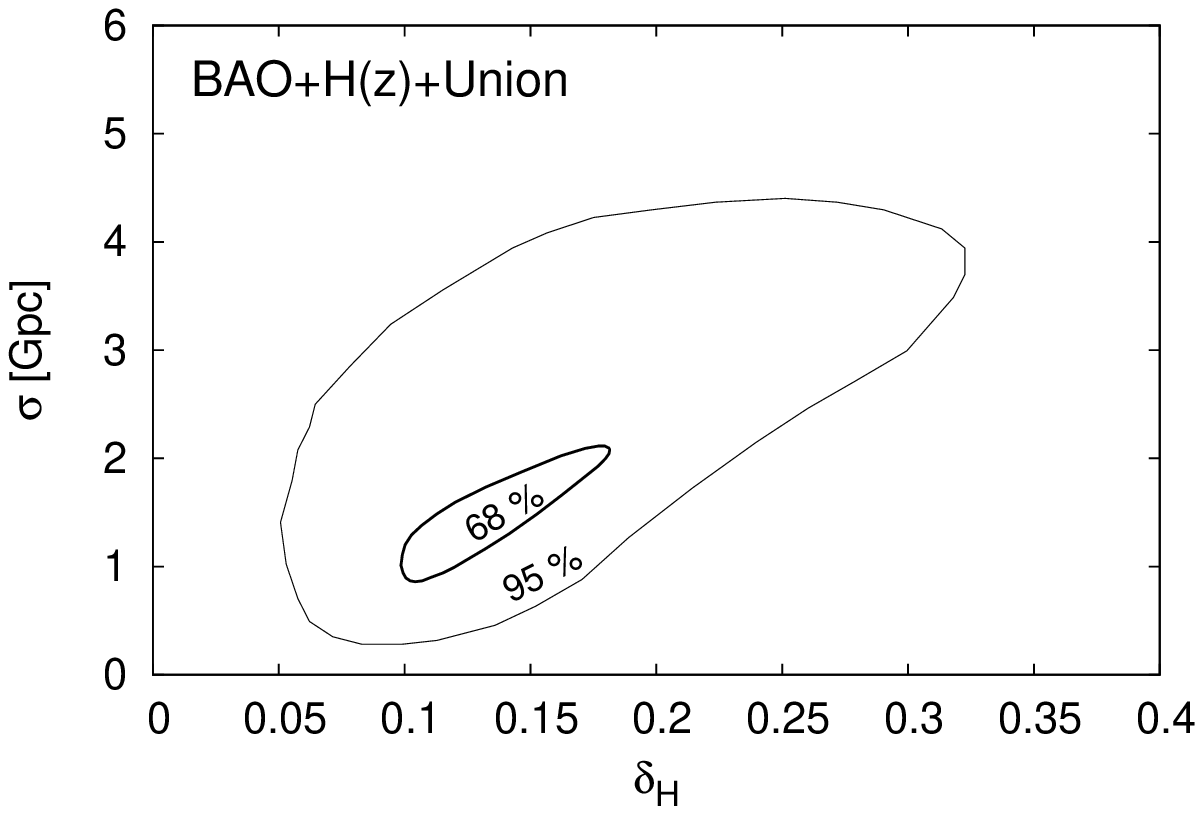}
\includegraphics[scale=0.4]{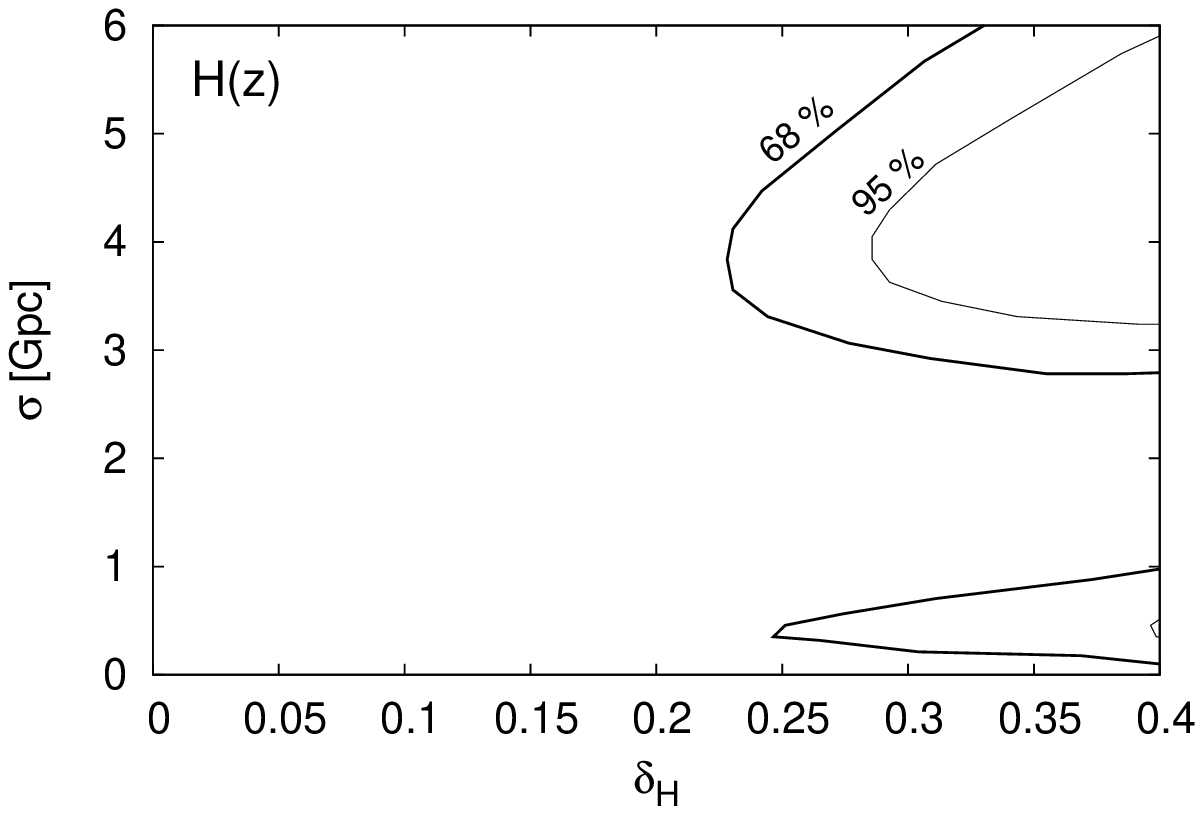}
\includegraphics[scale=0.4]{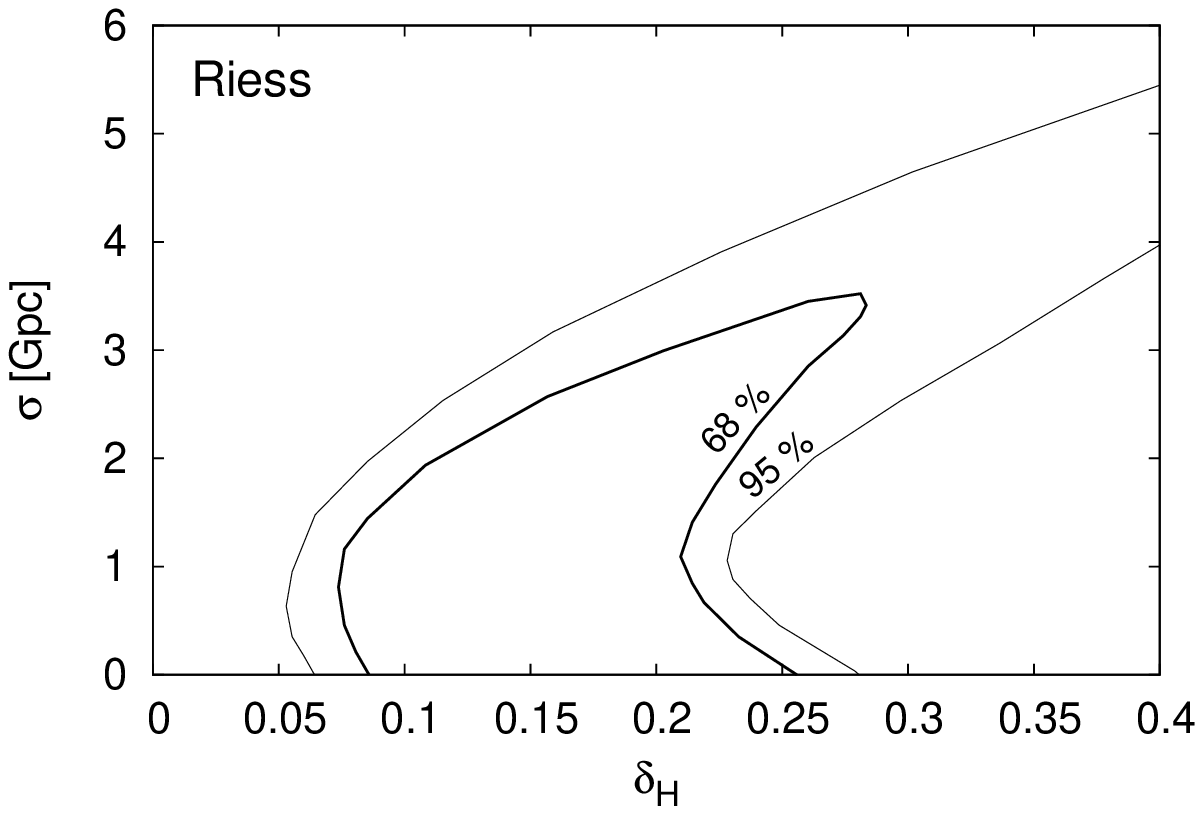}
\includegraphics[scale=0.4]{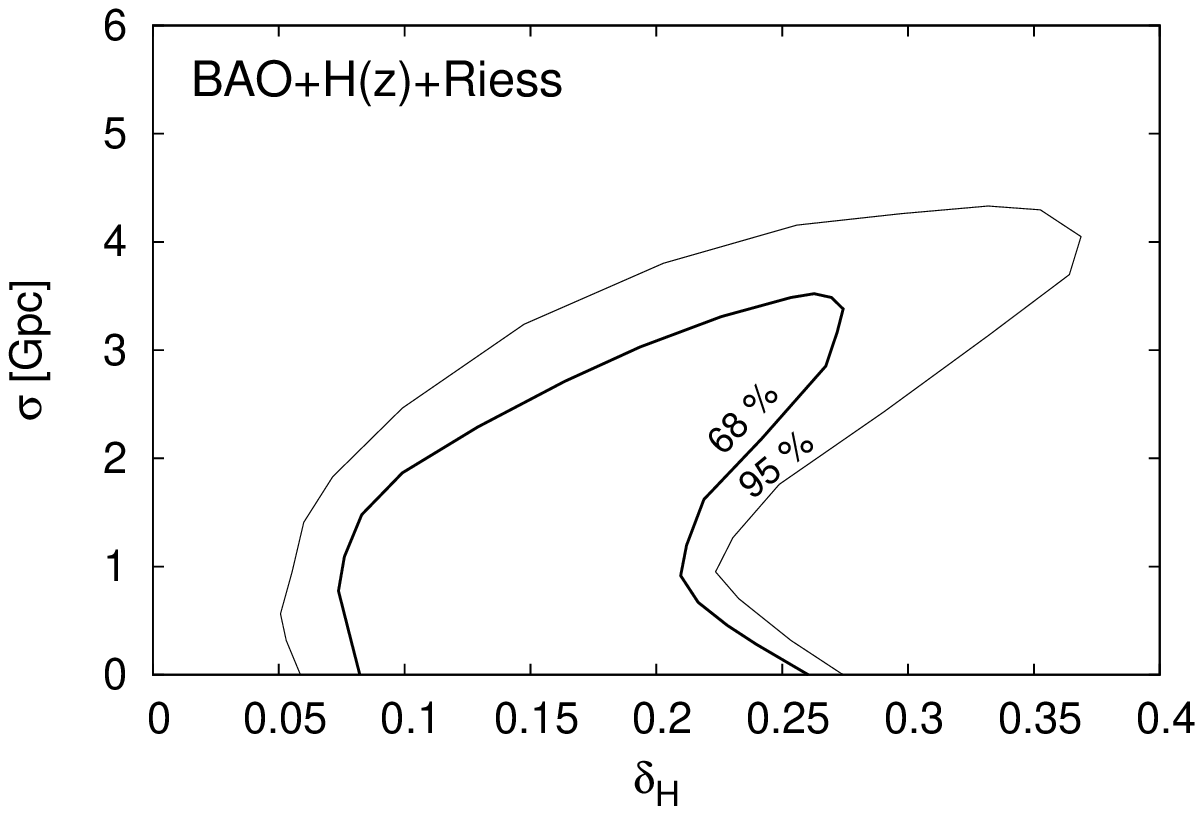}
\caption{Observational constraints on the parameters of the second family of models (cosmic flow models), shown as contour plots of $\chi^2$ for the model given the data,
after marginalizing over $H_0$.
The confidence levels are set assuming a $\chi^2$ distribution.
{\em Upper Left}: constraints from the BAO.
{\em Lower Left}: constraints from H(z).
{\em Upper Center}: constraints from the Union data set.
{\em Lower Center}: constraints from the Riess gold data set.
{\em Upper Right}: joint constraints from 
the BAO+H(z)+Union data sets.
{\em Lower Right}: joint constraints from 
the BAO+H(z)+Riess data sets.}
\label{fig5}
\end{center}
\end{figure}

As above we allow three parameters to vary in the range $H_0 \in [64,76]$ km s$^{-1}$ Mpc$^{-1}$, $\sigma \in [0,6]$ Gpc, and $\delta_H \in [0, 0.4]$. The likelihood distributions for these parameters,
based on the BAO+H(z)+Union data sets are presented in Fig.~\ref{fig4}.
The constraints on the parameters 
$\delta_H$ and $\sigma$ are 
shown as contour plots of $\chi^2$ after marginalizing over $H_0$ in Fig.~\ref{fig5}.
This family of models provides an even better fit to available observations. The best-fit model fits the data (BAO+H(z)+Union)
with $\chi^2 =317.64$ (for 314 degrees of freedom)
which means that the model can be ruled out at only the 56.8\% level.
The best-fit model has $H_0 \delta_H \approx 8.5$ km s$^{-1}$ Mpc$^{-1}$ and $\sigma \approx 1.3$ Gpc.
We note that this large scale flow is not compatible with the original Hubble Bubble from Ref. \cite{JRK07}  where $H_0 \delta_H \approx 4.5$ km s$^{-1}$ Mpc$^{-1}$ and $\sigma \approx 0.1$ Gpc. Indeed, as seen from Fig.~\ref{fig5} 
the Hubble Bubble can be ruled out at $2 \sigma$ confidence.
However, when the Riess gold data set is used 
to constrain the model, 
the Hubble Bubble is only ruled out at the 1-sigma level.

\begin{table}
\centering
\caption{Expected values (means) for $H_0$, $\sigma$, and $\delta_{\rho}$
based on the BAO+H(z)+Union data sets.}
\footnotesize\rm
\begin{tabular}{lccc}
\br
model  & $H_0$  & $\sigma$ & $\delta$ \\
 & [km s$^{-1}$ Mpc$^{-1}$] & [Gpc] & \\
\mr 
cosmic void  & $67.62^{+0.49}_{-0.48}$  & $2.96^{+0.19}_{-0.45}$ & $4.05^{+1.28}_{-1.44}$\\
cosmic flow  & $69.24^{+0.66}_{-0.62}$  & $1.46^{+0.55}_{-0.26}$ & $0.13^{+0.02}_{-0.01}$ \\
\br
\label{tab1}
\end{tabular}
\end{table}

\begin{table}
\centering
\caption{Parameters and the $\chi^2$ of the best-fit models
based on the BAO+H(z)+Union data sets.}
\footnotesize\rm
\begin{tabular}{lccccc}
\br
model  & $H_0$  & $\sigma$ & $\delta$ & $\chi^2$ & $\chi^2/{\rm dof}$ \\
 & [km s$^{-1}$ Mpc$^{-1}$] & [Gpc] & & & \\
\mr 
cosmic void  & $67.603 $  & $3.119$ & $4.501$ & 336.55 & 1.07\\
cosmic flow  & $69.209$  & $1.327$ & $0.123$ & 317.64 & 1.01\\
\br
\label{tab2}
\end{tabular}
\end{table}

\section{Future observational constraints}\label{FPsec}

\begin{figure}
\begin{center}
\includegraphics[scale=0.33, angle=270]{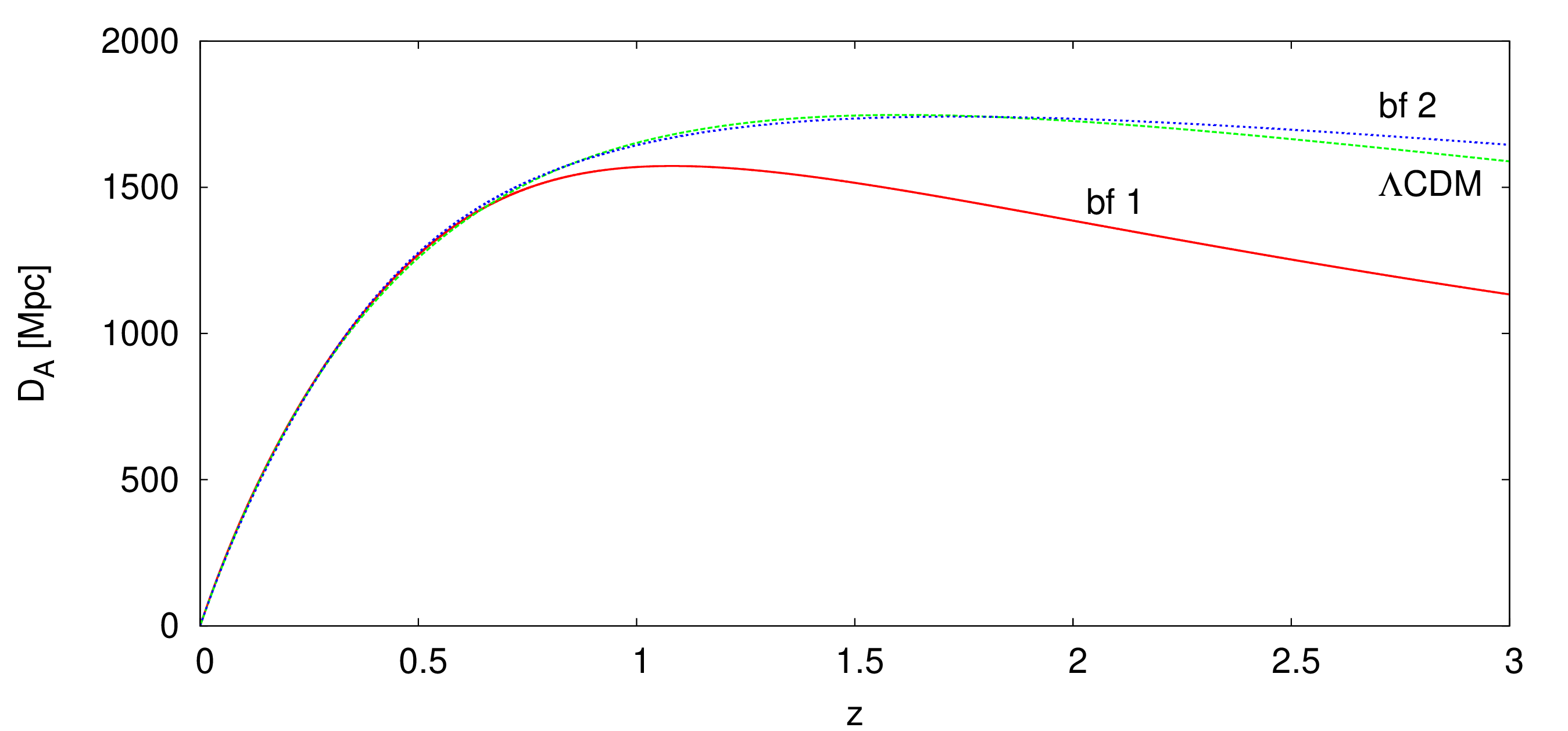}
\caption{The angular diameter distance as a function of redshift in the ${\Lambda}$CDM model and the best-fit inhomogeneous models 
(bf 1 -- best-fit cosmic void model, bf 2 -- best-fit cosmic flow model).
The inset presents the absolute distance difference, i.e. 
$(D-D_{{\Lambda}{\rm CDM}})/D$ (where $D$ is the angular distance
either in bf 1 or bf 2 model).}
\label{fig6}
\end{center}
\end{figure}

\begin{figure}
\begin{center}
\includegraphics[scale=0.8]{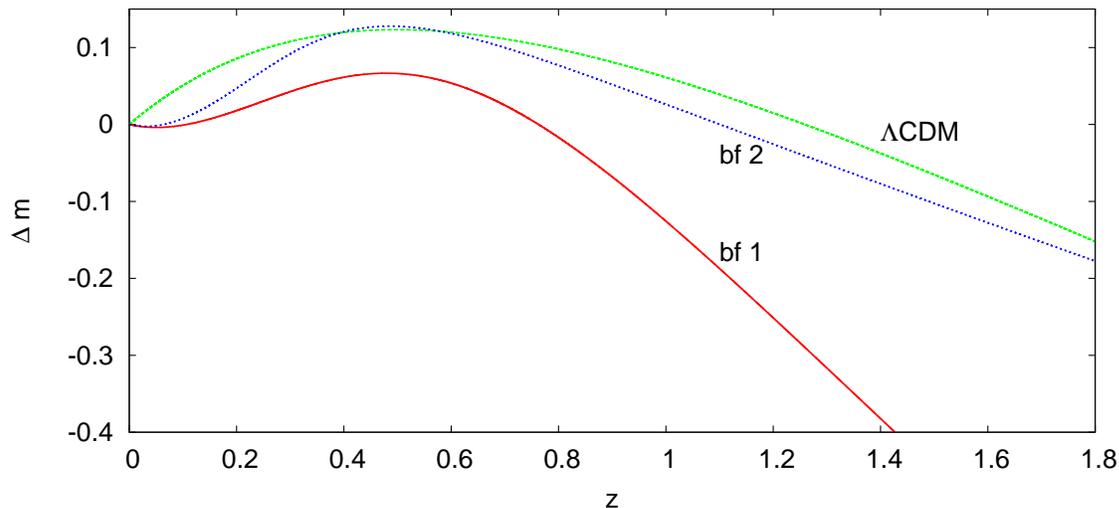}
\caption{The residual Hubble diagrams for the best-fit void model
(bf 1), best-fit cosmic flow model (bf 2), and the ${\Lambda}$CDM model.}
\label{fig7}
\end{center}
\end{figure}

The previous section showed that very large regions of parameter space 
for inhomogeneous models can be ruled out by combinations of
existing constraints. However, a large scale void model 
 or an ultra-large cosmic flow remain as possible
alternatives to dark energy with respect to explaining the apparent
acceleration of the Universe. 
In our analysis we did not consider constraints from the CMB.
This is because observations at high redshift ($z\gg1$) are of limited
significance when used to constrain the properties of the local-scale inhomogeneities.
Owing to the flexibility of inhomogeneous models 
we can always assume that our void is surrounded by an overdense or underdense region --
there is not need to assume that the Universe consists only of the void
of Gpc-radius and everywhere else is homogeneous.
Thus, by varying the density of such a ring or its size, every model studied in this paper can be made to fit the distance to the 
last scattering surface (which is given by the position of the first of the CMB peak).
Thus, the high redshift data does not significantly constrain the 
properties of a local inhomogeneity (see Appendix for more details). To tighten the constraints on the properties
of the local Gpc-scale inhomogeneity we need
more precise data sets at
low redshift which can be directly compared with constraints from
existing data sets such as Type-Ia supernovae. Among these observations,
the most valuable are those of distance, of the Hubble parameter, or measurements like baryonic acoustic oscillations and the Alcock--Paczy\'nski
tests which
depend on both the Hubble parameter and distance. 
For example, precise measurements of a maximum in the angular diameter distance distance would place very tight constraints \cite{H06,AS07}.
This is depicted in Fig.~\ref{fig6} which
shows the angular distance 
as a function of redshift in the ${\Lambda}$CDM model and in the best fitted models for both families of models
studied in this paper. The angular distance to  $z>1$ in inhomogeneous void models  is distinguishable from the corresponding distance in homogeneous models, at a level larger then 10\%. In addition 
and the position of maximum for model bf 1 is shifted by $\Delta z \approx 0.5$ 
relative to the $\Lambda$CDM model. However,
current estimates of the angular distance from SnIa are predominantly at $z<1$, and are not sufficiently precise to discriminate between models on this basis. Other state of the art the measurements of the distance based on the Sunyaev-Zel'dovich effect are also too imprecise~\cite{B06}.
Another possible test that has been proposed in the literature is 
based on the time drift of cosmological redshift, i.e. a measurement of $\dot{z}$ \cite{UCE08}.
However, this requires a very precise measurement since the typical amplitude
of the effect  is $\delta z \sim 10^{-10}$ on a time scale
of 10 years  \cite{UCE08}.
Another possible test was proposed in \cite{CBL08}, which 
relies on studying the consistency relation between the distance on the null cone and 
the Hubble parameter. Such relation does not hold if the geometry of the Universe
differs from the FLRW geometry.
 Of more immediate utility are methods based on 
spectral distortions of the CMB power spectrum \cite{G95}.
These have already been used to put an upper bound of 
 $\approx$ 1 Gpc for the radius of a local Gpc-scale void \cite{CS08}.
In addition, a study of the kinematic Sunyaev--Zel'dovich effect has  recently been used
by Garc\'ia-Bellido and Haugb\o lle \cite{GH08b}
to estimate the properties of a local void. This analysis shows that the void
permitted by the existing data cannot be larger than $\sim 1.5$ Gpc.
We recall from Sec. \ref{sec5.1} that a void  with $\sigma = 1$ Gpc and $\delta_{\rho} = 0.95$ 
fits the data with $\chi^2 = 347.27$, which means that 
that the model can be rejected at only the $90.5\%$ level, which 
is insufficient to rule out this
possibility. Moreover, cosmic-flow models, 
although not tested with these type of observations 
(kSZ effect and spectral distortion), require an inhomogeneity
of radius also around 1 Gpc.
Therefore, there is a need for future observational constraints
to either confirm or rule out the existence of these large inhomogeneities.

Of considerable promise for the future  are very precise measurements of supernova in the redshift range of 0.1-0.4 \cite{CFL08}. In this redshift range (see Fig.~\ref{fig7})
measurements of over 2000 supernova (for example by the Joint Dark Energy Mission) will be 
sufficient to reassure existence of such a void.
However, in the near future  the most promising  candidate for testing inhomogeneous cosmological models are BAO measurements. 
Current measurements of BAO from LRG SDSS (for an up to date status see \cite{GCH08}) 
already add additional constrains (Figs.~\ref{fig3} and \ref{fig5}).
Moreover, the WiggleZ project \footnote{http://wigglez.swin.edu.au}
will soon measure  the dilation distance with accuracy of $2.5\%$ at $z = 0.75$.
In the next subsection we show that this forthcoming observation
will be sufficient to significantly constrain the inhomogeneous cosmological models.

\subsection{BAO constraints}\label{sec6.1}

Before proceeding with further analysis, 
we asses the importance of analysing BAO data using 
inhomogeneous, rather than FLRW models.
For example, when analysing the data in \cite{E05}
the redshifts of LRG galaxies were translated into comoving coordinates 
using the flat FLRW model with $\Omega_m=0.3$, $\Omega_{\Lambda} =0.7$,
$H_0 = 70$ km s$^{-1}$ Mpc$^{-1}$.
However, if the $\Lambda$CDM model is not a good background
model of the Universe then such a procedure leads to inaccurate results.
The error which arises from this conversion
can be estimated from the inset in Fig.~\ref{fig6}.
As seen at $z =0.35$, the error in the distance conversion 
is $3\%$ and $0.3\%$ respectively 
in cases where the best-fit model from 
the first (bf 1) and second (bf 2) family of models studied in this paper 
were chosen as the background rather than the $\Lambda$CDM model.
These results indicate that while the effect is relatively small,
we should be aware that the value $D_V$ at $z=0.35$ estimated in \cite{E05}
when using the inhomogeneous model could be different than $D_V=1370 \pm 64$ Mpc.
On the other hand we find that even if this value was change by $5\%$ the results presented in the previous 
section would still hold. This is because the, as seen from Figs. \ref{fig3}
and \ref{fig5} the observations of BAO at $z=0.35$ do not put tight constraints 
on the inhomogeneous models relative to  data.
However at higher redshift ($z\approx 1$) the redshift-distance conversion error
increases to around $10\%$ for the cosmic void model.
Thus, when using forthcoming BAO data to constrain inhomogeneous models
one must perform with the whole analyses within the inhomogeneous 
framework.

In this section we will assume for the sake of argument 
 that such an analysis has been performed.
We will also assume that the $D_V$ is equal to the value obtained within the $\Lambda$CDM model ($\Omega_m=0.3$, $\Omega_{\Lambda} =0.7$,
$H_0 = 70$ km s$^{-1}$ Mpc$^{-1}$).
As in Sec. \ref{sec5} we set $\sigma \in [0,6] Gpc$,
$\delta_{\rho} \in [0,6]$, and $\delta_H \in [0,0.4]$.
However, we tighten $H_0$ range to $H_0 \in [68,72]$,
because future observational experiments like
WiggleZ project will measure the Hubble parameter with $4\%$
precision.

\begin{figure}
\begin{center}
\includegraphics[scale=0.55]{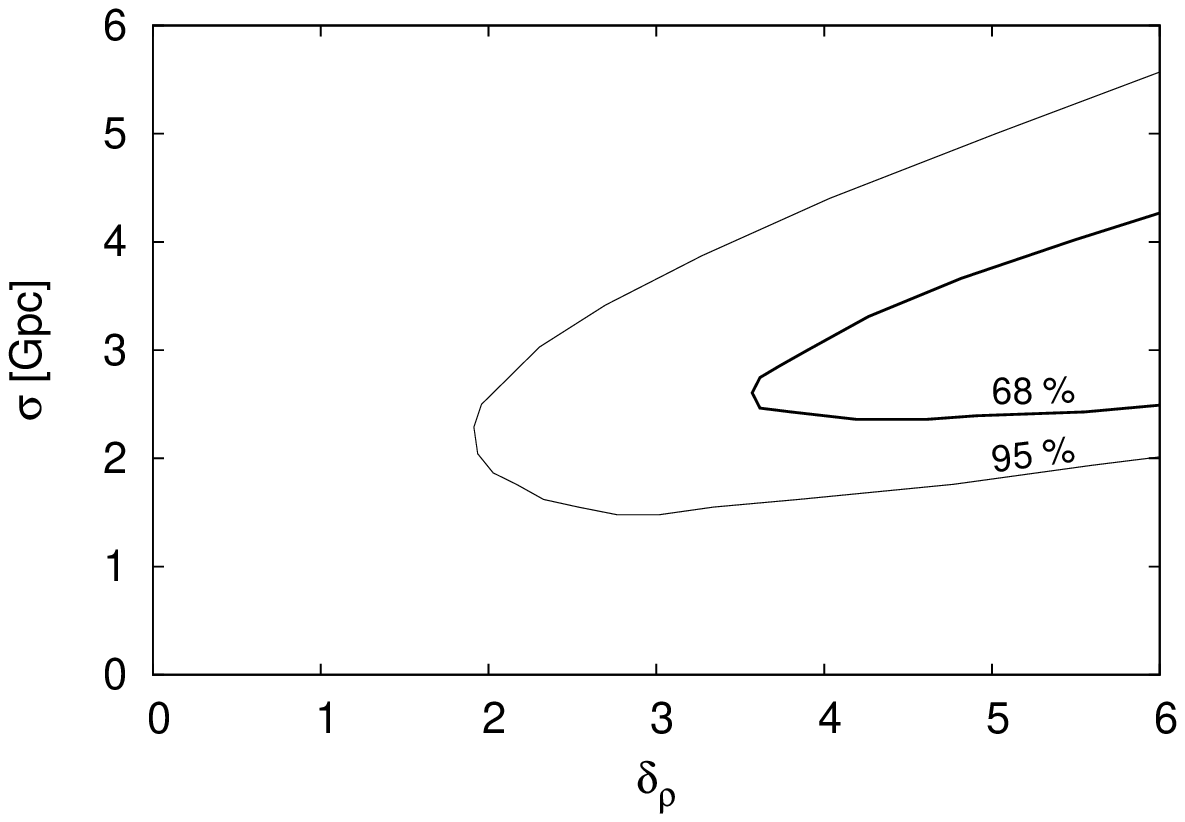}
\includegraphics[scale=0.55]{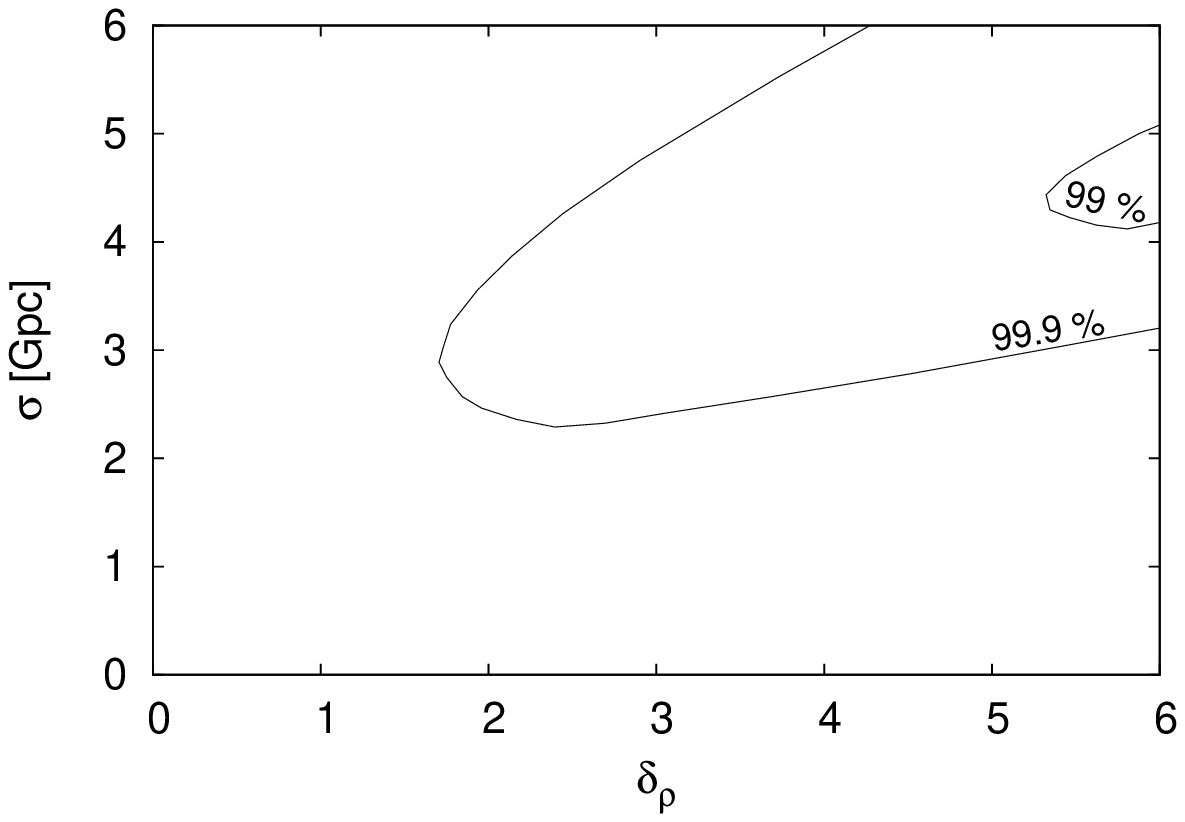}
\includegraphics[scale=0.55]{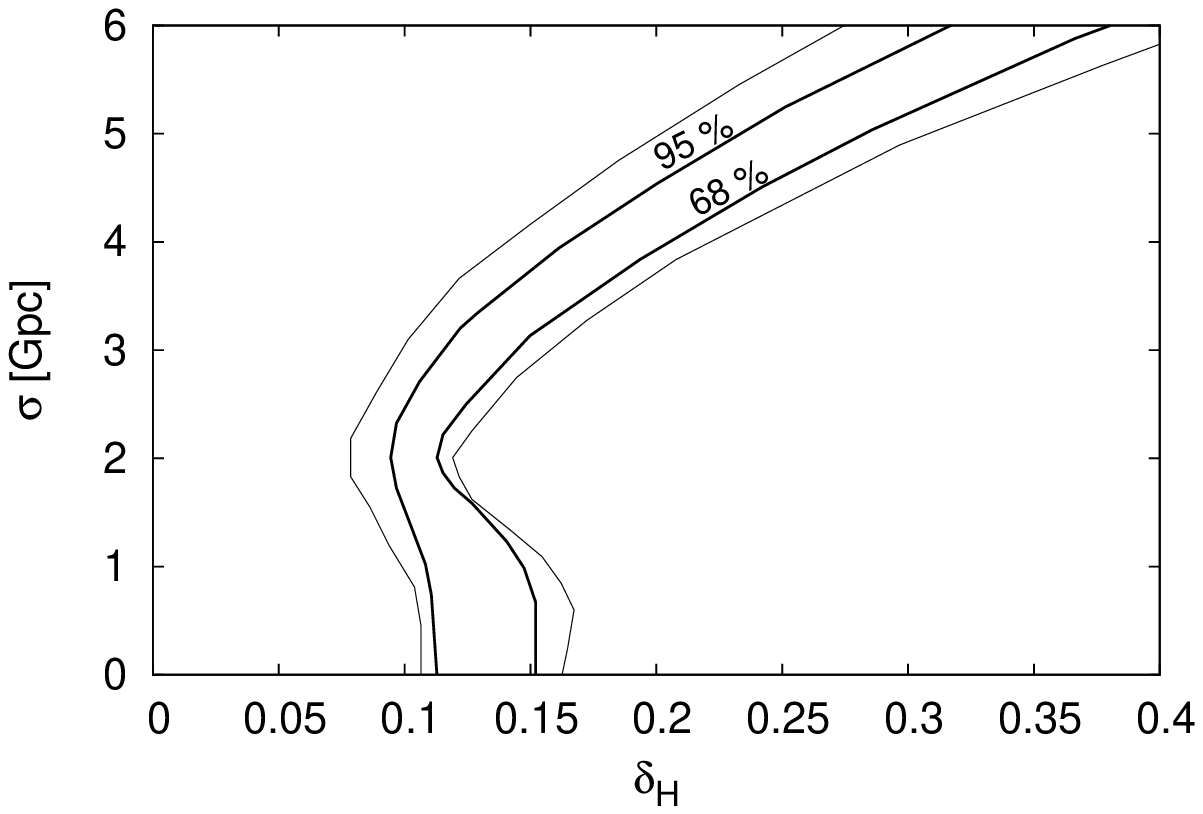}
\includegraphics[scale=0.55]{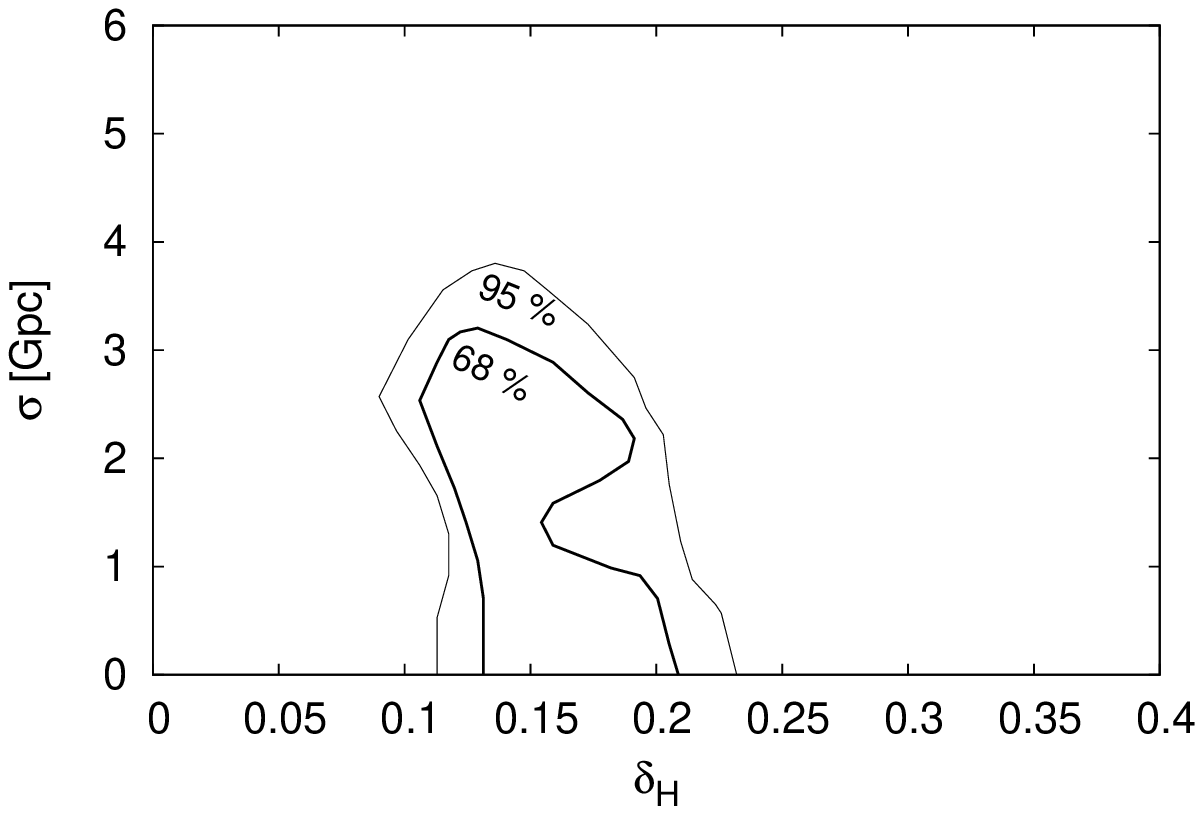}
\caption{ 
Observational constraints on the parameters of the first family of models, shown
as contour plots of $\chi^2$ for the model given the data,
after marginalizing over $H_0$.
 The confidence levels are set assuming that the value of $D_V$ is the same as in the ${\rm \Lambda}$CDM model and that it is measured with 2.5\% precision.
{\em Upper Left}: expected constraints from the BAO at z=0.75 for the first family of models (cosmic void).
{\em Upper Right}: expected constraints from the BAO at z=0.95 for the first family of models (cosmic void).
{\em Lower Left}: expected constraints from the BAO at z=0.75 for the second family of models (cosmic flow).
{\em Lower Right}: expected constraints from the 
8 BAO measurements at z=0.35,0.55,0.75,...,1.75.}
\label{fig8}
\end{center}
\end{figure}

The potential constraints from this BAO experiment around the 
$\Lambda$CDM model are presented in the left panels of Fig.~\ref{fig8}.
As measurement of the BAO scale at $z=0.75$ with a precision of $2.5\%$
would rule out most of the parameter space considered for the cosmic depression  family of models. A measurement at $z=0.95$ with $2.5\%$ accuracy  would be sufficient to rule out  all remaining possibilities in this case at a significance of $3 \sigma$.

The second family of models considered in this paper 
are not tightly constrained by a single BAO measurement at $z=0.75$.
As seen in lower right panel of Fig.~\ref{fig8} even measurements up to redshift 1.75 are not able to rule
out this type of model. 
To rule out all remaining possibilities at a significance of $3 \sigma$  measurements should be extended beyond $z=2$.

\section{Conclusions}

Homogeneous cosmological models 
which include a dark energy component and which are 
built on the Copernican Principle have been spectacularly successful in describing a suite of cosmological observations.
However, alternatives which replace dark energy with an 
gigaparsec-scale inhomogeneity 
have also been postulated as 
explanations for the apparent acceleration of the Universe.
In this paper we have shown that most inhomogeneous anti-Copernican models can already be ruled out by combining current cosmological observations (SNIa, H(z), and BAO).
Our model were based on a simple parametrization of an inhomogeneity.
We assumed that the inhomogeneity is modeled with a Gaussian-like
profile, although other profiles are also possible.
For example a similar analysis was recently proposed by Garc\'ia-Bellido and Haugb\o lle \cite{GH08c}
were the inhomogeneity was modeled using a hyperbolic tangent function.
Their results also support our conclusion that 
 there is still a range of parameter space that remains consistent with current constraints. 
As a result the success of Concordance models cannot be considered as a validation of the Copernican Principle.
Forthcoming experiments 
will be capable of constraining all possible alternatives and will therefore 
either validate the Copernican Principle or determine the existence of a local Gpc-scale inhomogeneity.

\ack
This research was supported by the Peter and Patricia Gruber Foundation
and the International Astronomical Union (KB) and by
the Australian Research Council (JSBW).
We thank Aleksandra Kurek for helpful discussions.

\appendix

\section{Fitting the CMB data with inhomogeneous models}

\setcounter{section}{1}

In the standard approach, the CMB temperature fluctuations are analyzed 
by solving the Boltzmann equation
within linear perturbation around the 
homogeneous and isotropic FLRW model \cite{SZ96,SSWZ03}\footnote{This approach is implemented in such
codes like CMBFAST (http://www.cfa.harvard.edu/$\sim$mzaldarr/CMBFAST/cmbfast.html),
 CAMB (http://www.camb.info/),
or CMBEASY (http://www.cmbeasy.org/).}.
Within an inhomogeneous background one can proceed with a similar analysis  -- employing 
the LT model instead of the FLRW model. Alternatively, if it is assumed that the early Universe 
(before and up to the last scattering instant) is well described by the FLRW model
then the CMB power spectrum can be parametrized by \cite{DL02}

\begin{equation}
l_m = l_a (m - \phi_m)
\end{equation}
where $l_1, l_2, l_3$ is a position of the first, second and third peak, and

\begin{equation}
l_a = \pi \frac{R}{r_s},
\end{equation}
where $R$ is a comoving distance to the last scattering surface and $r_s$ is a size of the
sound horizon at the last scattering instant.
The sound horizon depends on $\Omega_b,~\Omega_m,~\Omega_{\gamma}$, and $h$
(where
$\Omega_b,~\Omega_m,~\Omega_{\gamma}$ are ratios of baryon, mater and radiation 
energy density respectively to the critical energy density evaluated at the current instant
and $h$ is equal to H/100 km s$^{-1}$ Mpc$^{-1}$). The size of the sound horizon can be calculated accordingly to formulae given in \cite{EH98}.
The function $\phi_m$ depends on $\Omega_b,~\Omega_m,~h,~n_s$, and the  
energy density of dark energy ($n_s$ is the spectral index). The exact form of 
$\phi_m$ is given in \cite{DL02}.
The $\Omega_m$ and $h$ are given by a ratio 
$\rho/ \rho_{cr}$ and $H/(100 {\rm~km~s^{-1} ~Mpc^{-1}})$ at the last scattering surface
(evaluated for the current instant).

In this appendix we will present a fit to the CMB data using the above formula.
Let us consider four models

\begin{itemize} 
\item model A1:

$ \rho(t_0,r) = \rho_b \left[ 1 + 1.45 - 1.45\exp \left( - \frac{\ell^2}{0.75^2} \right) \right], $

$  t_B = 0, \quad \Omega_b = 0.0445, \quad n_s = 1, $

where $\ell = r/Gpc$.

\item model A1+ring:

$ \rho(t_0,r) = \rho_b \left[ 1 + 1.45 - 1.45\exp \left( - \frac{\ell^2}{0.75^2} \right) 
-1.75 \exp \left\{ - \left( \frac{\ell- 5.64}{0.926} \right)^2 \right\} 
\right], $

$  t_B = 0, \quad \Omega_b = 0.08, \quad n_s = 0.963. $

\item model A2

$ \rho(t_0,r) = \rho_b \left[ 3.33  + 1.4 \exp \left( - \frac{\ell^2}{0.75^2} \right) \right], $

$  t_B = 0, \quad \Omega_b = 0.0445, \quad n_s = 1. $

\item model A2+ring:

$ \rho(t_0,r) = \rho_b \left[ 
 3.33  + 1.4 \exp \left( - \frac{\ell^2}{0.75^2} \right)
+ 0.08 \exp \left\{ - \left( \frac{\ell- 7.5}{0.965} \right)^2 \right\} 
\right], $

$  t_B = 0, \quad \Omega_b = 0.07, \quad n_s = 0.963. $

\end{itemize} 

\begin{figure}
\begin{center}
\includegraphics[scale=0.6]{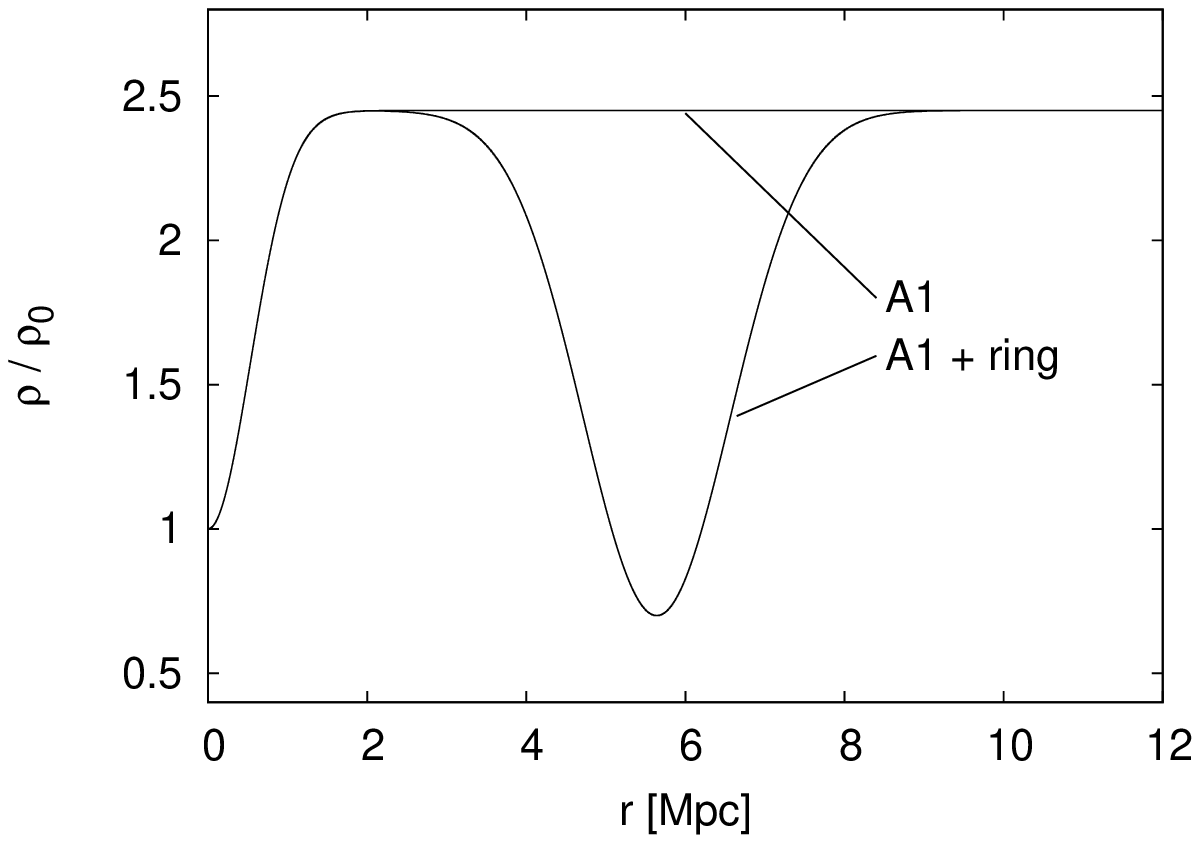}
\includegraphics[scale=0.6]{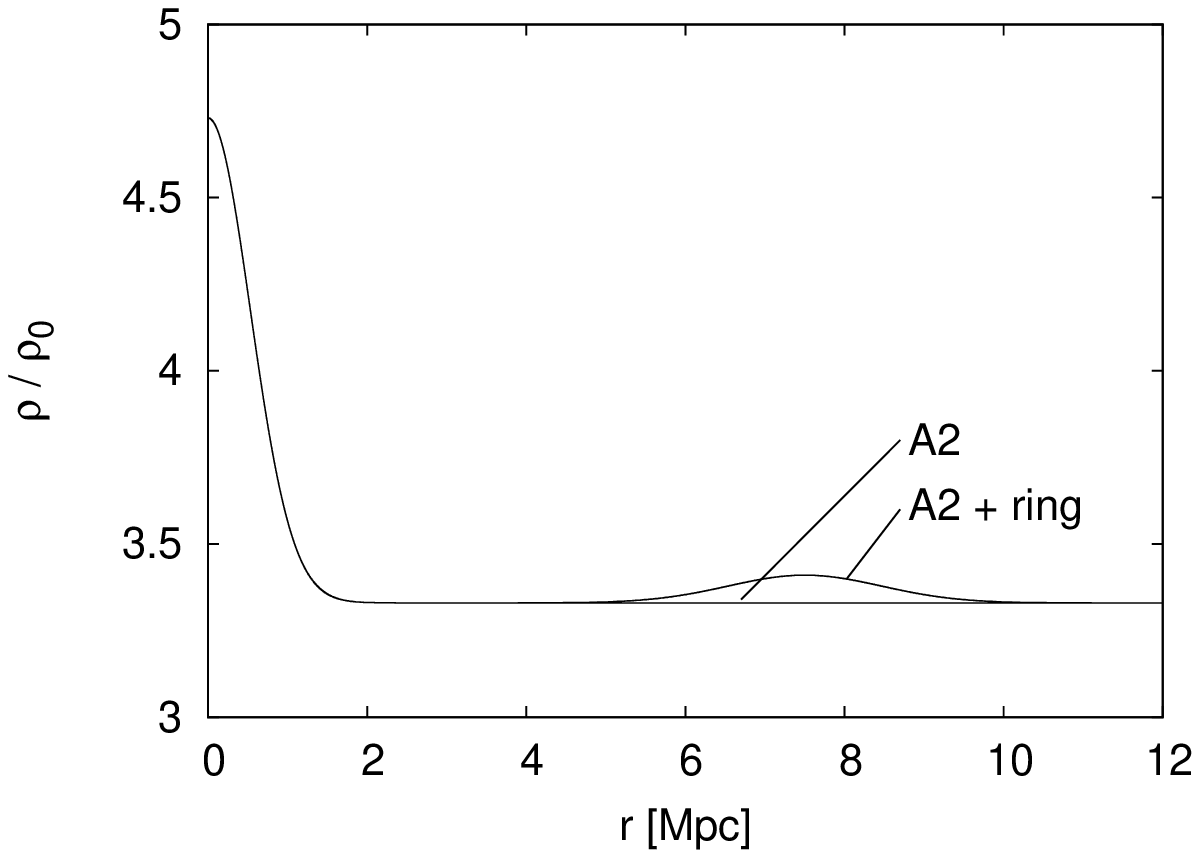}
\caption{Density distribution at the current instant for models A1, A2, A1+ring, A2+ring.}
\label{fa1}
\end{center}
\end{figure}

The density distributions of these models are presented in Fig.~\ref{fa1}.
The fit to positions of the CMB peaks is presented in Table \ref{Ta1}.
As can be seen, model A1 does not fit the observed CMB power spectrum.
However, we can modify the density distribution is such a way (by adding an underdense ring)
that the CMB peaks can be reproduced  -- model A1+ring.
In models where density increases up to some distance from the origin
(local void type of models) a satisfactory fit to the first peak of CMB power spectrum
(i.e. distance to the last scattering instant) can be obtained if the mass of the universe is decreased.
This decrease can be obtained either by having a local void of a larger radius (as in \cite{GH08a} where $R \approx 2.5$ Gpc),
or by having an addition underdense ring between local void and the last scattering surface.

On the other hand, it is not only the mass that is important. 
To show this let us consider model A2. In model A2 the density decreases up to some distance from the origin.
In this configuration we can also obtain a good 
fit to the position of the first peak if we increase the mass of the universe.

The positions of other peaks strongly depend on $\Omega_b$.
To obtain a good fit to the positions of other peaks 
in our model we need to increase  the value of $\Omega_b$ 
beyond the value that is consistent with observation
of light elements in the local Universe ($\Omega_b = 0.0445$ \cite{C04}) to 
$\Omega_b = 0.08$. However, due to the lower expansion rate
at large distances, the physical baryon density $\Omega_b h^2$ remains 
is close to the observed value, i.e. $0.0198$  and $0.0212$ for models
A1+ring and A2+ring respectively.

The above results suggest that almost every model can be modified in such a way that 
a good fit to CMB power spectrum data can be obtained. 
Models A1 and A2 are very different and yet after some modifications they are both able to fit the CMB data. Thus, the CMB data
does not strongly constrain the properties of the local Gpc void.
Each model for the local scale void considered in this paper 
can be modify in such a way (by adding one or two rings between a local void and the last scattering surface) so that the CMB power spectrum is recovered.

\begin{table}
\centering
\label{Ta1}
\caption{CMB fit}
\footnotesize\rm
\begin{tabular}{lccc}
\br
Model & First peak & Second peak & Third peak \\
\mr
A1 & 177.34 & 414.08 & 627.19 \\
A2 & 216.59 & 503.37 & 754.28 \\
A1 + ring & 220.46 & 530.38 & 800.19 \\
A2 + ring & 220.43 & 530.51 & 779.76 \\
\mr
WMAP \cite{WMAP3,WMAP5} & $220.8 \pm 0.7$  & $530.9 \pm 3.8$ & $700-1000$ \\
\br
\end{tabular}
\end{table}

\end{document}